\documentclass{article}

\usepackage{amssymb}
\usepackage{tikz}
\usetikzlibrary{shapes}
\usepackage[applemac]{inputenc}
\usepackage{color}

\usepackage{marvosym}
\usepackage{algorithm}
\usepackage{algorithmic}
\usepackage{calc}
\usepackage{tikz}

\usetikzlibrary{positioning}
\usetikzlibrary{backgrounds}
\usetikzlibrary{patterns}

\usepackage{fullpage}

\usepackage{eufrak}
\usepackage{ifthen}
\usepackage{amsthm}
\usepackage{amsmath}

\newtheorem{theorem}{Theorem}
\newtheorem{lemma}{Lemma}
\newtheorem{corollary}{Corollary}
\newtheorem{remark}{Remark}

\newtheorem{example}{Example}

\title{Flow methods for cooperative games with generalized coalition configuration}

\author{Encarnaci\'on Algaba\footnote{Matem\'atica Aplicada II and Instituto de Matem\'aticas de la Universidad de Sevilla (IMUS), Escuela  Superior de Ingenieros, Camino 
de los Descubrimientos, s/n, 41092 Sevilla, Spain.  ealgaba@us.es} 
\and Eric R\'emila \footnote{Universit\'e de Saint-Etienne, CNRS, GATE Lyon-St-Etienne UMR 5824, F-42023 Saint-Etienne, France, eric.remila@univ-st-etienne.fr}  \and Philippe Solal \footnote{Universit\'e de Saint-Etienne, CNRS, GATE Lyon-St-Etienne UMR 5824, F-42023 Saint-Etienne, France,  philippe.solal@univ-st-etienne.fr
 } }

\date{}

\begin{document}

\maketitle

\abstract{This paper introduces the class of cooperative games with generalized coalition configuration. This new class of games corresponds to cooperative games with coalition configuration and restricted cooperation. A coalition configuration is a collection of coalitions covering the agent set. The restriction of cooperation between agents is represented by a set system on each element of the coalition configuration. A coalition profile is a list of feasible coalitions, one for each element of the coalition configuration. A coalition profile function associates a worth with each coalition profile. Based on this framework, we define and axiomatically characterize marginal values whose coefficients induce a unitary flow on the product digraph obtained from these set systems. Next, we propose a two-step procedure, inspired by Owen's procedure, to construct flow methods as above. Then, we show that the associated flow is decomposable into two flows. Finally, we use two axioms to characterize the flows that can be decomposed in this way, and hence the flow methods constructed using our procedure.}



\section{Introduction}
Given a finite set of agents, a coalition configuration is a collection of coalitions on that set such that the union of these coalitions is equal to this set of agents (Albizuri et al. \cite{Albizuri2006a}).
We consider cooperative games with coalition configuration, with the novelty that for each element of the coalition configuration, a set system represents the set of coalitions that can be formed within this element, thus generalizing  coalition configuration structures considered so far in the literature.
These structures are referred to as generalized coalition configurations. In this setting, a coalition profile is defined as a list of feasible coalitions, one for each element of the coalition configuration. This profile represents the way in which the population organizes itself to cooperate in each element of the coalition configuration. The coalition profile function associates a worth to each feasible coalition profile.
A value for these cooperative games with generalized coalition configuration is a function that associates a payoff vector to each game of this class. As a first step, we consider values that satisfy three principles inspired by the axioms for cooperative games with transferable utility (in short TU-games): Efficiency, Linearity and the Null agent principle. We show that these three principles induce marginalist values whose coefficients lead to a unitary flow on the product digraph constructed from the cartesian product of the covering relations of the set systems (Theorem \ref{th:marg_cov} and Theorem \ref{flow-Eff}). Such a value is called a flow method. Theorem \ref{th:marg_cov} and Theorem \ref{flow-Eff} extend 
Theorem 2.4 and Theorem 2.10 in Aguilera et al. \cite{Aguilera2010} obtained for TU-games with restricted cooperation. Flow methods are allocation methods popularized by Weber \cite{Weber1988} for TU-games and by Moulin and Vohra \cite{Moulin-2003} for cost-sharing problems.

In a second step, we construct Owen-type values \cite{Owen1977} using a two-step procedure. In the first step of the procedure, the elements of the coalition configuration play a game with no restrictions on cooperation. In this step, a flow method is applied. The flow passes through the edges of the directed hypercube formed by all the coalition profiles that can be formed between the elements of the coalition configuration. In a second step of the procedure, the agents of each element of the coalition configuration play a cooperative game that takes into account both the cooperation restrictions in that element of the coalition configuration and the result of the previous step. A flow method is applied at this step of the procedure. The flow passes through the edges of the directed graph associated with the covering relation of the set system of this element of the coalition configuration. We obtain the following results.

Firstly, Theorem \ref{char-flowmethod} shows that Owen-type values obtained by this procedure are flow methods, called two-step flow methods. The flow goes through the directed edges on the product digraph defined above. Moreover, this flow can be decomposed into two flows. More precisely, it can be calculated as the product of the flow resulting from the first step of the procedure and the flow resulting from the second step of the procedure. Among the family of such two-step flow methods are those based on maximal paths, in the spirit of Shapley's procedure \cite{Shapley1953}, where the grand coalition is formed step by step from a linear order on the agents set. With the same idea Owen \cite{Owen1977} and Albizuri et al. \cite{Albizuri2006a} construct their value from certain maximal paths and unitary flows on these paths. This allows us to extend the configuration value of Albizuri et al. \cite{Albizuri2006a}{\footnote{Albizuri and Aurrekoetxea \cite{Albizuri2006b} also provide a generalization of the normalized Banzhaf-Coleman index (Dubey and Shapley \cite{DS1979}) in the setting of cooperative games with coalition configuration. Likewise, Albizuri and Vidal-Puga \cite{Albizuri2015} study weighted bounded configuration values.} to our more general framework where the unit of cooperation is not a coalition but a coalition profile and where the cooperation is restricted in each element of the coalition configuration. 

A second result focuses on the characterization of two-step flow methods among the flow methods. Theorem \ref{char-twostepflow1} and Theorem \ref{char-twostepflow} provide an axiomatic characterization of two-step flow methods using two axioms about flows. The first axiom states that certain directed edges of the product digraph receive no flows. The second axiom is a principle of flow proportionality between certain directed edges of the product digraph. Next, we return to the situation where cooperation among the agents is not restricted. Among the flow methods,  we point out a configuration-type value for our situation where the unit of cooperation are not coalitions but coalition profiles. Theorem  \ref{AZ-theo} provides an axiomatic characterization of this configuration-type value among the flow methods in terms of two axioms of anonymity for flows and the axiom defined above concerning null flow on certain directed edges of the product digraph.

Finally, we explore the possibility of moving from coalition profiles to coalitions. More precisely, we address the question of the conditions under which a flow method defined for situations where the unit of cooperation is a coalition profile induces a flow method for situations where the unit of cooperation is the coalition calculated from the union of the coalitions in a coalition profile. Theorem  \ref{final} provides such a condition on the product digraph, which is satisfied if the coalition configuration is a partition of the agent set or if each set system is regular. It should be noted that by flow method here, we mean a marginalist and efficient value whose coefficients define a flow on a digraph built from the set of feasible coalitions obtained from the feasible coalition profiles. This digraph
 does not necessarily coincide with the digraph of the covering relation over the set of feasible coalitions, but it is consistent with the underlying product digraph. Nervertheless, it coincides with the digraph of the covering relation in the two cases above: the coalition configuration is a partition of the agent set or if each set system is regular. 
 
The rest of the article is organized as follows. Section \ref{Prelim} provides general definitions about set systems, digraphs and flows. Section \ref{Section-Games} introduces cooperative games with generalized coalition configuration, i.e., cooperative games with coalition configuration and restricted cooperation.
Section \ref{Sec-Marg} introduces marginalist values and flow methods for this class of cooperative games. Section \ref{Sec-Marg-Charact}  provides an axiomatic characterization of the marginalist values and flow methods. Section \ref{Section-Two-step-FM} introduces and characterizes two-step flow methods. Section \ref{Section-Two-step-FM-Chara} proposes two axioms for flows and shows that these axioms single out the flow of the two-step flow methods. Section \ref{coalitionp-coalition} discusses the conditions under which a flow method for games where the relevant unit of cooperation is a coalition profile induces a flow method for games where the relevant unit of cooperation is reduced to a coalition. Section \ref{concl} concludes.

\section{Preliminaries} \label{Prelim}

\subsection{Set systems and digraphs}

Let $N \subseteq \mathbb{N}$ be a finite set of $n\geq 2$ agents, and let $2^N$ be the collection of all subsets of $N$. Each element $K$ of $2^N$ is interpreted as a {\bf coalition} of agents that sign a binding agreement.
A {\bf set system} over $N$ is a pair $(N, {\cal F})$, where ${\cal F} \subseteq 2^N$ is a collection of coalitions of $N$. 
A set system is {\bf normal} if $\{\emptyset, N\} \subseteq {\cal F}$.  For any two feasible coalitions $K$ and $K'$ of ${\cal F}$, $K'$ {\bf covers} $K$ if $K \subset K'$ and $K'' \in {\cal F}$ such that $K \subseteq K'' \subset K'$ implies $K'' = K$. Define the associated {\bf (acyclic) directed graph} (or just {\bf digraph})  $\Gamma_{\cal F} = ({\cal F}, E_{{\cal F}})$ where  $(K, K') \in E_{\cal F}$ if $K' $ covers $K$; $K$ and $K'$ are the {\bf endpoints} of  $(K, K')$, $K$ is the {\bf tail} of $(K, K')$ and $K'$ is its {\bf head}. 
A {\bf (directed) path} on  $\Gamma_{\cal F}$ is an ordered sequence of distinct vertices (coalitions) $W = (K_0, K_1, \ldots, K_q)$, $q \geq 1$, such that $(K_r, K_{r+1})\in E_{\cal{F}}$ for each $r \in \{0, \ldots, q -1\}$.
The {\bf length} of the path  $W= (K_0, K_1, \ldots, K_q)$ is $q$. A path $W= (K_0, K_1, \ldots, K_q)$ is {\bf maximal} if $K_0 = \emptyset$ and $K_q = N$.
A {\bf quasi-path} on  $\Gamma_{\cal F}$ is an ordered sequence of distinct vertices (coalitions) $W = (K_0, K_1, \ldots, K_q)$, $q \geq 1$ such that either $(K_r, K_{r+1})\in E_{{\cal F}}$ or  $(K_{r+1}, K_{r})\in E_{{\cal F}}$ for each $r \in \{0, \ldots, q-1\}$. A {\bf component} of $\Gamma_{\cal F}$ is a maximal subset (with respect to set inclusion)  of  ${\cal F}$ such that each pair of its elements can be connected by a quasi-path.
The set of components of a digraph $\Gamma$ is denoted by  ${\cal C}_{\Gamma}$.

\subsection{Flows}

Given a normal set system and its associated directed graph $\Gamma_{\cal F} = ({\cal F}, E_{\cal F}) $ a {\bf flow} on  $\Gamma_{\cal{F}}$ is a mapping $\Lambda: E_{\cal F} \longrightarrow \mathbb{R}$ that satisfies the following {\bf conservation constraints}:
$$ \forall K \not \in \{\emptyset, N\}, \quad  \sum_{(K', K) \in E_{\cal F} }\Lambda (K', K) = \sum_{(K, K'') \in E_{\cal F} }\Lambda (K, K'').$$
The {\bf value} $V(\Lambda)$ of a flow $\Lambda $ on  $\Gamma_{\cal F}$  is given by

$$V(\Lambda) = \sum_{(\emptyset, K) \in E_{\cal F}} \Lambda (\emptyset, K)  =  \sum_{(K, N) \in E_{\cal F}} \Lambda (K, N). $$ 
 A flow $\Lambda $ is {\bf unitary} if  $V(\Lambda) = 1$. \\

A {\bf cut} of $\Gamma_{\cal F} $ is a partition of ${\cal F}$ formed by two subsets  ${\cal F}_{\emptyset}$ and ${\cal F}_{N}$ such that  $\emptyset \in {\cal F}_{\emptyset}$ and $N \in {\cal F}_{N}$. Indifferently, a cut $({\cal F}_{\emptyset}, {\cal F}_{N})$ is sometimes seen  as  the set of  directed edges of $E_{\cal F} $ with an endpoint  in ${\cal F}_{\emptyset}$ and  the other endpoint in ${\cal F}_{N}$.
The {\bf value of a cut} $({\cal F}_{\emptyset}, {\cal F}_{N}  )$ is defined as the net flow across the cut:
$$\sum_{(K', K) \in E_{\cal F} \cap ({\cal F}_{\emptyset} \times {\cal F}_{N} ) }\Lambda (K', K) - \sum_{(K, K') \in E_{\cal F} \cap ({\cal F}_{N} \times {\cal F}_{\emptyset} )}\Lambda (K, K').$$

It is known that the value $V(\Lambda)$ of a flow $\Lambda$ is equal to the value of any cut $({\cal F}_{\emptyset}, {\cal F}_{N}  )$   (see, e.g., Gross and Yellen \cite{Gross2006}, Proposition 13-1-3):

\begin{equation}
V(\Lambda) = \sum_{(K', K) \in E_{\cal F} \cap ({\cal F}_{\emptyset} \times {\cal F}_{N} ) }\Lambda (K', K) - \sum_{(K, K') \in E_{\cal F} \cap ({\cal F}_{N} \times {\cal F}_{\emptyset} )}\Lambda (K, K').
\end{equation}

\section{Cooperative games with generalized coalition configuration} \label{Section-Games}

Let $N = \{1, 2, \ldots, n \}$ be a finite agent set of size $n \in \mathbb{N}$. Every subset of $N$ is called a {\bf coalition}. 
Following Albizuri et al. \cite{Albizuri2006a}, a {\bf coalition configuration} is a list $\mathcal {P} = (P_1, P_2, \ldots, P_m)$, $m\geq 1$, of non-empty coalitions that cover $N$, i.e.,
$$\bigcup_{q = 1}^{m} P_q = N.$$
These structures generalizes the so-called coalition structures introduced by Aumann and Dr\`eze \cite{AD1974}. A {\bf coalition structure} is a coalition configuration  $\mathcal {P}$ that forms a partition of $N$.

However, the assumption of full cooperation within each coalition of a coalition configuration is not always realistic, as there may be asymmetries or incompatibilities between agents in each coalition of a coalition configuration.
This is why we introduce the notion of {\bf generalized coalition configuration} which takes into account the possibilities of restricted cooperation in each element of the coalition configuration.
Hence, denote by $M = \{1, \ldots, m \}$ the index set whose elements label the elements of $\mathcal {P}$. For each $q \in M$,  $(P_q, {\cal F}_q)$ represents a normal set system.
Each element of ${\cal F}_q$ is a {\bf feasible coalition} of $P_q$. Denote by ${\cal F} $ the cartesian product 
 $$ \prod_{q = 1}^{m} {\cal F}_q.$$
A {\bf feasible coalition profile} is an element ${\cal K} = (K_1, \ldots, K_m) $ of  ${\cal F}$, and $\emptyset_M$ denote the empty coalition profile $(\emptyset, \ldots, \emptyset )$. 
For each ${\cal K}$ and $q \in M$,  ${\cal K}_{-q}$ denotes the restricted profile without component $q$, i.e., 
$${\cal K}_{-q}= (K_1, \ldots, K_{q-1}, K_{q+1},\ldots, K_m).$$
The {\bf support} $\mu ({\cal K})$ of a  feasible coalition profile ${\cal K} = (K_1, \ldots, K_m)$ is the subset of $M$ defined as:
$$ \mu ({\cal K}) = \bigl\{q \in M: K_q \not = \emptyset\bigr\}.$$
The set ${\cal F}$ can be ordered through the product relation $ \subseteq^M$:   
$$\forall {\cal K}, {\cal K}' \in {\cal F}, \quad {\cal K} \subseteq^M {\cal K} \,\, \mbox{ if } \,\, \forall q \in M, \,\, K_q \subseteq K'_q.$$
We say that the coalition profile ${\cal K'}$ covers ${\cal K}$ with respect to $ \subseteq^M$ if there is a unique $q \in M$ such that
${\cal K}'_{-q} = {\cal K}_{-q}$ and $(K_q, K'_q) \in E_{{\cal F}_q}$. 
Denote by $q_{({\cal K}, {\cal K}')}$ just this component of the vector, and let $\Gamma_{{\cal F}} = ({\cal F}, E_{{\cal F}})$ be the associated {\bf digraph of the covering relation} of $({\cal F}, \subseteq^M)$, which is the cartesian product of the digraphs $\Gamma_{{\cal F}_q}= ({\cal F}_q, E_{{\cal F}_q} ) $ associated with the covering relation of the set systems $(P_q, {\cal F}_q)$, $q \in M$. For each directed edge $({\cal K}, {\cal K}') \in  E_{{\cal F}}$, the subset $Q_{({\cal K}, {\cal K}')}$ stands for the subset of agents in $K'_{q_{({\cal K}, {\cal K}')}}\setminus K_{q_{({\cal K}, {\cal K}')}}$.
For each agent $i \in N$,  $E_{{\cal F}}^i \subseteq E_{{\cal F}}$ denotes the subset of directed edges in which $i$ is involved:
$$E_{{\cal F}}^i =\bigl \{({\cal K}, {\cal K}') \in  E_{{\cal F}}: i \in Q_{({\cal K}, {\cal K}')} \bigr\}, $$
that is, it represents the set of situations where $i$ and possibly other agents enter a coalition of an element $P_q$, for some $q \in M$, of the coalition configuration ${\cal P}$, knowing that this agent may already be present in a coalition of another element $P_{q'}$, $q' \in M \setminus q$, of the coalition configuration ${\cal P}$. This allows us to define for each $i \in N$, the {\bf subdigraph} $\Gamma_{{\cal F}}^i$ induced by the directed edges $E_{{\cal F}}^i$. The endpoints of the elements of $E_{{\cal F}}^i$ forms the vertex (coalition) set $V_{\cal F}^i  \subseteq {\cal F}$.   In general, $\Gamma_{{\cal F}}^i = (V_{\cal F}^i , E_{\cal F}^i)$  is not a bipartite digraph. If the coalition configuration ${\cal P}$ is a partition, i.e., a coalition structure, then $\Gamma_{{\cal F}}^i$ is bipartite for each $i \in N$.  The following example illustrates the above definitions.

\begin{example} \label{graph1}
Assume that  $N = \{1, 2, 3, 4, 5\}$, ${\cal P} = \{ P_1, P_2\}$ where $P_1 = \{1, 2, 3\}$ and $P_2 = \{3, 4, 5\}$. The subset of feasible coalitions in $P_1$ and in $P_2$ respectively are given by the following set systems:
$${\cal F}_1 = \{\emptyset, \{1\}, \{2, 3\}, \{1, 2, 3\}\} \mbox { and } {\cal F}_2 = \{\emptyset, \{3, 4\},  \{3, 4, 5\}\}.$$ Figure \ref{fig1} represents the digraphs $\Gamma_{{\cal F}_{1}} = ({{\cal F}_1}, E_{{\cal F}_1})$ and $\Gamma_{{\cal F}_{2}} = ({{\cal F}_2}, E_{{\cal F}_2})$; Figure \ref{fig2-1} represents the product digraph $\Gamma_{{\cal F}_{}} = ({{\cal F}}, E_{{\cal F}})$ and the digraph $\Gamma_{{\cal F}}^3$ corresponding to Agent 3, respectively. Note that
 the directed edge   $$((\emptyset, \{3, 4\}), (\{2,3\}, \{3, 4\})),$$ of  $\Gamma_{{\cal F}}^3$ represents a situation where agent 3 and agent 2 form a coalition in $P_1$ while agent 3 is already cooperating with agent 4 in $P_2 $.
 \end{example}
 
 \begin{figure}[h]
\begin{center}
\begin{tikzpicture}  [every node/.style={scale = 0.6}, scale = 0.6]
 \node (A) at (0,0) {$\emptyset$};
  \node (B) at (5,0) {$\{2, 3\}$};f
  \node (C) at (2.5, 1.5) {$\{1\}$};
  \node (D) at (7.5, 1.5) {$\{1, 2, 3\}$};

  \draw[->] (A) -- (B);
  \draw[->] (B) -- (D);
  \draw [->] (C) -- (D);
  \draw [->] (A) -- (C);

%
%
  \node (E) at (12,0) {$\emptyset$};
  \node (F) at (12, 2) {$\{ 3, 4\}$};
  \node (G) at (12,4) {$\{3, 4, 5\}$};
  \draw[->] (E) -- (F);
  \draw[->] (F) -- (G);
\end{tikzpicture}
\end{center}
\caption{The digraphs $\Gamma_{{\cal F}_{1}} = ({{\cal F}_1}, E_{{\cal F}_1})$ and $\Gamma_{{\cal F}_{2}} = ({{\cal F}_2}, E_{{\cal F}_2})$} 
\label{fig1}
\end{figure}
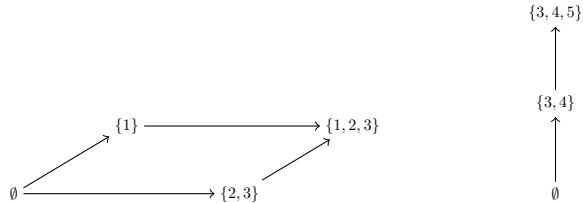

\begin{figure}[h]
\begin{center}
\begin{tikzpicture} [every node/.style={scale = 0.6}, scale = 0.6]
  \node (A) at (0,0) {$(\emptyset, \emptyset)$};
  \node (B) at (5,0) {$(\{2, 3\}, \emptyset)$};
  \node (C) at (2.5, 1.5) {$(\{1\}, \emptyset)$};
  \node (D) at (7.5, 1.5) {$(\{1, 2, 3\}, \emptyset)$};

  \draw[->] (A) -- (B);
  \draw[->] (B) -- (D);
  \draw [->] (C) -- (D);
  \draw [->] (A) -- (C);

 \node (E) at (0, 3) {$(\emptyset, \{3, 4\})$};
 \node (F) at (5, 3) {$(\{2, 3\}, \{3, 4\})$};
\node (G) at (2.5 , 4.5) {$(\{1\}, \{3, 4\})$};
\node (H) at (7.5, 4.5) {$(\small \{1, 2, 3\}, \{3, 4\})$};  
  
   \draw[->] (E) -- (F);
  \draw[->] (E) -- (G);
  \draw [->] (G) -- (H);
  \draw [->] (F) -- (H);
  
  \node (I) at (0, 6) {$(\emptyset, \{3, 4, 5\})$};
\node (J) at (5, 6) {(\{2, 3\}, \{3, 4, 5\})};
\node (K) at (2.5, 7.5) {(\{1\}, \{3, 4, 5\})};
\node (L) at (7.5, 7.5) {(\{1, 2, 3\}, \{3, 4, 5\})};

   \draw [->] (I) -- (J);
 \draw [->] (I) -- (K);
 \draw [->] (J) -- (L);
 \draw [->] (K) -- (L);  

\draw [->] (A) -- (E);
\draw [->] (E) -- (I);

  \draw [->] (B) -- (F);
  \draw [->] (F) -- (J);
  
  \draw [->] (C) -- (G);
  \draw [->] (G) -- (K);  
  
    \draw [->] (D) -- (H);
  \draw [->] (H) -- (L);  
  
  
    \node (A) at (10,0) {$(\emptyset, \emptyset)$};
  \node (B) at (15,0) {$(\{2, 3\}, \emptyset)$};
  \node (C) at (12.5, 1.5) {$(\{1\}, \emptyset)$};
  \node (D) at (17.5, 1.5) {$(\{1, 2, 3\}, \emptyset)$};

  \draw[->] (A) -- (B);
  \draw [->] (C) -- (D);

 \node (E) at (10, 3) {$(\emptyset, \{3, 4\})$};
 \node (F) at (15, 3) {$(\{2, 3\}, \{3, 4\})$};
\node (G) at (12.5 , 4.5) {$(\{1\}, \{3, 4\})$};
\node (H) at (17.5, 4.5) {$(\small \{1, 2, 3\}, \{3, 4\})$};  
  
   \draw[->] (E) -- (F);
  \draw [->] (G) -- (H);
  
\node (I) at (10, 6) {$(\emptyset, \{3, 4, 5\})$};
\node (J) at (15, 6) {(\{2, 3\}, \{3, 4, 5\})};
\node (K) at (12.5, 7.5) {(\{1\}, \{3, 4, 5\})};
\node (L) at (17.5, 7.5) {(\{1, 2, 3\}, \{3, 4, 5\})};

   \draw [->] (I) -- (J);
 \draw [->] (K) -- (L);  

\draw [->] (A) -- (E);

  \draw [->] (B) -- (F);
  
  \draw [->] (C) -- (G);
  
    \draw [->] (D) -- (H);
 
  \end{tikzpicture}
\end{center}
\caption{To the left, the product digraph $\Gamma_{{\cal F}} = ({\cal F}, E_{{\cal F}})$. To the right, the subdigraph $\Gamma_{{\cal F}}^3 = (V_{\cal F}^3 , E_{\cal F}^3)$ with four components.}
\label{fig2-1}
\end{figure}
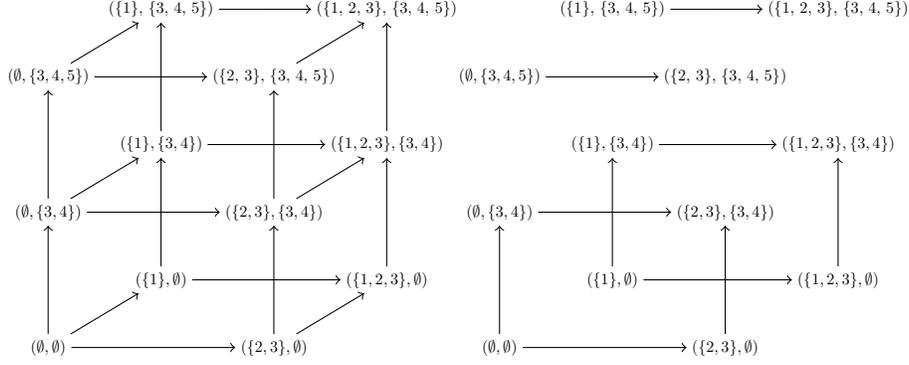

Given a product digraph $\Gamma_{{\cal F}} = ({\cal F}, E_{{\cal F}})$,
a {\bf coalition profile function} $v: {\cal F} \longrightarrow \mathbb{R}$ assigns a worth $v({\cal K}) \in  \mathbb{R}$ to each ${\cal K} \in  {\cal F}$, where, by convention,  $v(\emptyset_M) = 0 $. 
This describes a situation in which agents organized themselves into a family of coalitions not necessarily pairwise disjoint (the coalition configuration). Inside each of these coalitions the cooperation is restricted so that only feasible coalition profiles can form.
If such a feasible coalition profile forms then its members can cooperate by signing an agreement in each feasible coalition of this profile, and then earn the worth associated with this profile.
Given a coalition profile ${\cal K}  \in  {\cal F} \setminus \emptyset_M$, the {\bf Dirac function} $1_{{\cal K}}: {\cal F} \longrightarrow \mathbb{R}$
is defined as
$$1_{{\cal K}} ( {\cal K}') = 1  \mbox { if } {\cal K}' = {\cal K}  \quad \mbox { and }  \quad   1_{{\cal K}} ( {\cal K}') = 0    \mbox{ otherwise. }              $$
Obviously, $v: {\cal F} \longrightarrow \mathbb{R}$ admits the linear decomposition
\begin{equation} \label{Dirac}
v = \sum_{{\cal K} \in {\cal F} \setminus \emptyset_M} v({\cal K})1_{\cal K} .
\end{equation}

\begin{remark} \label{r1} In case the coalition configuration is formed only by the grand coalition $N$, then the situation is identical to a cooperative TU-game with restricted cooperation as studied 
by Aguilera et al. \cite{Aguilera2010}.
In the latter case, if the set system is formed by all possible coalitions of $N$, then the situation is a classical TU-game. This also appears when $P_i = \{i\}
$ for each $i \in N$. The above framework can also be viewed as a 
way to model externalities across coalitions: the worth of a coalition $K_q \in {\cal F}_q$ depends on the organization ${\cal K}_{-q}$ in  $(P_1, \ldots, P_{q-1}, P_{q+1},\ldots, P_m)$.
Nevertheless, the way the above framework models externalities is different from both the model of cooperative games in configuration form with externalities in Albizuri  \cite{Albizuri2010} and the model of cooperative games with restricted cooperation and externalities in Albizuri et al. \cite{Albizuri2023}. In these models there is no a priori coalition configuration from which coalition profiles can form. These two models rely on the concept of embedded coalition that specifies a feasible coalition $K$ together with a feasible coalition configuration ${\cal K}$ (i.e., the coalitions of ${\cal K}$ necessarily covers $N$). The function $v$ is defined over the set of embedded coalitions of the form $(K, {\cal K})$ so that $v( K, {\cal K})$ can be different from $v( K', {\cal K})$ for $K$ and $K'$ in ${\cal K}$, which cannot be the case in our framework. In Albizuri et al. \cite{Albizuri2023}, ${\cal K}$ is supposed to be a partition of the agent set, and all singleton coalitions are feasible. In Albizuri \cite{Albizuri2010}, all coalition configurations are feasible except those which contain comparable coalitions with respect to set inclusion.    
\end{remark}

A {\bf cooperative game with generalized coalition configuration} is a tuple $$(N, v, \mathcal {P}, \mathcal {F}),$$
where $N$ is a finite set of agents, $\mathcal {P} = (P_1, \dots, P_m) $ is an a priori coalition configuration, $\mathcal {F} $ represents the set of feasible coalition profiles that can be formed from cartesian product of the normal set systems $(P_q, {\cal F}_q)$, $q \in M$, and $v$ is a coalition profile function defined on $\mathcal {F}$. In what follows, it is assumed that $N$, ${\cal P}$ and ${\cal F}$ are fixed once and for all, so that only the coalition profile function $v$ can vary. From now on, we denote by   ${\cal G}_{N, {\cal P}, {\cal F}}$ 
the domain of cooperative games with generalized coalition configuration.

\section{Marginalist values and flow methods} \label{Sec-Marg}

A {\bf  payoff vector} for a cooperative game with generalized coalition configuration $(N, v, \mathcal {P}, \mathcal {F})$ is an $n$-dimensional vector $x \in \mathbb{R}^N$ assigning a payoff  $x_i \in \mathbb{R} $ to each agent $i \in N$ for its participation in the game. 
A {\bf value} on ${\cal G}_{N, {\cal P}, {\cal F}}$ is a function $\Phi$ which assigns a unique payoff vector
$\Phi (N, v, \mathcal {P}, \mathcal {F}) \in \mathbb{R}^N$ to each cooperative game with generalized coalition configuration $(N, v, \mathcal {P}, \mathcal {F}) \in {\cal G}_{N, {\cal P}, {\cal F}}$. A value  $\Phi$ is  {\bf marginalist}  if there exist functions 
$\lambda_{i}^{\Phi} : E_{\cal F}^i \longrightarrow \mathbb{R}$ defined for each $i \in N$ such that
 
 \begin{equation} \label{Marg}
 \Phi_i(N, v, {\cal P}, {\cal F}) = \sum_{({\cal K}, {\cal K}') \in E_{\cal F}^i} \lambda_{i}^{\Phi}({\cal K}, {\cal K}') \bigl(v( {\cal K}') - v({\cal K})\bigr).
\end{equation}

A marginalist value $\Phi$ is a {\bf flow method} on ${\cal G}_{N, {\cal P}, {\cal F}}$ if there is a function $\Lambda^{\Phi}_{\Gamma_{\cal F}}: E_{\cal F} \longrightarrow \mathbb{R}$ defined as
\begin{equation} \label{flowM}
\forall ({\cal K}, {\cal K}') \in E_{\cal F}, \quad  \Lambda_{}^{\Phi} ({\cal K}, {\cal K}') = \sum_{i \in Q_{  ({\cal K}, {\cal K}')} }  \lambda_{i}^{\Phi}({\cal K}, {\cal K}'),
\end{equation}
that induces a unitary flow on $\Gamma_{\cal F}$.

Note that if $\Phi$ is a flow method such that
\begin{equation} \label{IS} 
  \forall ({\cal K}, {\cal K}') \in  E_{\cal F},  \forall i, i' \in Q_{({\cal K}, {\cal K}')}, \quad     \lambda_{i}^{\Phi} ({\cal K}, {\cal K}')  = \lambda_{i'}^{\Phi} ({\cal K}, {\cal K}'), 
\end{equation}
then the flow method is completely determined by the induced flow $\Lambda_{\Gamma_{\cal F}}^{\Phi}$, that is:
$$\forall i \in N, \quad  \Phi_{i} (N, v, {\cal P}, {\cal F}) = \sum_{({\cal K}, {\cal K}') \in E_{\cal F}^i}  \frac{\Lambda_{}^{\Phi}({\cal K}, {\cal K}')}{| Q_{({\cal K}, {\cal K}')}|}  \bigl ( v({\cal K}') - v({\cal K})  \bigr).$$

As underlined in Remark \ref{r1}, Aguilera et al. \cite{Aguilera2010} consider TU-games with restricted cooperation on the form $(N, v, {\cal F})$ where $(N, {\cal F})$ is a normal set system on $N$ and $v: {\cal F} \longrightarrow \mathbb{R}$ is a set function which assigns a real worth to each feasible coalition. In this context, which corresponds to specific cooperative games with generalized coalition configuration where ${\cal P}$ is $\{N\}$,
the authors study the marginalist values and the flow methods.
 These solutions generalize the ones introduced by Bilbao \cite{Bilbao1998} and Bilbao and Edelman \cite{Bilbao2000} for TU-games on convex geometries, those of Bilbao and Ord\'o\~{n}ez \cite{Bilbao_Ordonez}  for normal augmenting systems and those of Lange and Grabisch  \cite{Lange2009} for regular TU-games.  
 
 A {\bf convex geometry} is (a) a normal set system,  (b) closed under intersection and (c) for each feasible coalition $S \in {\cal F}$ different from $N$ there is an $i \in N \setminus S$ such that  $S \cup i  \in {\cal F} $. 
 
 An {\bf augmenting system} is a set system (d) containing the empty set, (e) stable for union
 that is if $S$ and $T$ are in ${\cal F}$ and $S \cap T \not = \emptyset$, then $S \cup T \in {\cal F}$,\footnote{Property (e) generalizes the set of connected coalitions derived from a communication graph (Myerson \cite{Myerson1977}). Cooperative games augmented by a union stable set system  have been developed in Algaba et al. \cite{ALG1}, \cite{ALG2}.} 
 (f) for two  coalitions $S$, $T$ in ${\cal F}$  such that $S \subset T$, there is $i \in T \setminus S$ such that $S \cup i \in {\cal F}$. 
 
Properties (c) and (f) are augmenting/one-point extension properties. In convex geometries every maximal path is of length $n$. This is also the case for a normal augmenting system. An augmenting system is a convex geometry if and only if it is normal and closed under intersection (see Bilbao \cite{Bilbao2003}).

A  set system is {\bf regular} if (g) it is normal and (h) each of its maximal paths is of length $n$.

Every convex geometry (normal augmenting set system respectively) is a regular set system but not the other way around. The marginalist values and flow methods for TU-games with convex geometry, TU-games with normal augmenting set system and regular TU-games are themselves a generalization of the marginalist values and flow methods for TU-games with no restriction on cooperation studied by Weber \cite{Weber1988}. 

Following Weber \cite{Weber1988}, Bilbao   \cite{Bilbao1998}, Bilbao and Edelman   \cite{Bilbao2000}, Lange and Grabisch \cite{Lange2009}, Bilbao and  Ord\'o\~{n}ez   \cite{Bilbao_Ordonez}, and Aguilera et al. \cite{Aguilera2010} characterize the marginalist values by using two well-known principles: the linearity of the solution in $ v$ and the principle of null payoff for agents with null contributions.  To obtain the family of flow methods, it is added the standard principle of efficiency establishing that the worth of the grand coalition $N$ is fully redistributed to its members. In the next section, our purpose is to characterize marginalist values and flow methods using the same principles but adapting them to our broader context ${\cal G}_{N, {\cal P}, {\cal F}}$.

\section{Characterization of marginalist values and flow methods} \label{Sec-Marg-Charact}

We introduce the axioms of Linearity, Null agent, and Efficiency for a value on ${\cal G}_{N, {\cal P}, {\cal F}}$. Linearity
is a straightforward generalization of the same principle for TU-games. The principle of null agent needs to be adapted and interpreted in this setting. An agent $i \in N$ is called null in a game in ${\cal G}_{N, {\cal P}, {\cal F}}$ if the variation in worth induced by each pair of coalition profiles that are the endpoints of the directed edges $E_{\cal F}^i$ of $\Gamma_{\cal F}^i$ is null. In other words, each time $i$ joins, possibly with other agents, a feasible coalition of an element $P_q$ of ${\cal P}$ to which it belongs, the variation in worth is null. In this sense, the arrival of this agent, possibly with other agents, in any feasible coalition of each element $P_q$ to which it belongs adds no worth. The null agent axiom states that such a null agent  obtains a null payoff. Efficiency states that the worth of the coalition configuration ${\cal P}$ is fully redistributed to its members. \\

\noindent {\bf Linearity}. A value $\Phi$ on ${\cal G}_{N, {\cal P}, {\cal F}}$ is linear if for each $(N, v, {\cal P}, {\cal F})$ and $(N, v', {\cal P}, {\cal F})$ in ${\cal G}_{N, {\cal P}, {\cal F}}$ and $\alpha \in \mathbb{R}$, it holds that:
$$ \Phi (N, \alpha v + v', {\cal P}, {\cal F}) =  \alpha \Phi (N, v, {\cal P}, {\cal F}) + \Phi (N, v', {\cal P}, {\cal F}),$$ 
where the coalition profile functions $\alpha v$ and $v + v'$  are defined as $(\alpha v) ({\cal K}) = \alpha v({\cal K})$ and $ (v+ v') ({\cal K}) = v({\cal K}) + v'({\cal K}),$ for each feasible coalition profile $ {\cal K} \in {\cal F}$. \\

An agent is {\bf null} in $(N, v, {\cal P},{\cal F})$ if $v({\cal K}) = v({\cal K}')$ for each $({\cal K}, {\cal K'}) \in E_{\cal F}^i$. \\

\noindent {\bf Null agent axiom}. A value $\Phi$ on ${\cal G}_{N, {\cal P}, {\cal F}}$ satisfies the Null agent axiom if for each $(N, v, {\cal F}) \in {\cal G}_{N, {\cal P}, {\cal F}}$ and each null agent $i$ in  $(N, v, {\cal P}, {\cal F})$, it holds that:

$$\Phi_i (N, v, {\cal P}, {\cal F}) = 0.$$

\noindent {\bf Efficiency}. A value $\Phi$ on ${\cal G}_{N, {\cal P}, {\cal F}}$ is Efficient if for each $(N, v, {\cal P},{\cal F}) \in{\cal G}_{N, {\cal P}, {\cal F}}$, it holds that:
$$ \sum_{i \in N} \Phi_i (N, v, {\cal P}, {\cal F}) = v ({\cal P}).$$

\begin{theorem} \label{th:marg_cov} A value $\Phi$ on ${\cal G}_{N, {\cal P}, {\cal F}}$  is marginalist as defined in (\ref{Marg}) if and only if it satisfies Linearity and the Null agent axiom.
\end{theorem} 

The above result relies on the following lemma.

\begin{lemma} \label{lem:combi}
 Let  $\Gamma = (V, E)$  be an acyclic digraph with only one component.  
 Assume there is a function $x: V \longrightarrow \mathbb{R}$ that assigns a real number $x(j)$ to each vertex $j \in 
 V$ under the following constraint:
 $$\sum_{j \in V} x(j) = 0.$$
 The following assertions hold.
 \begin{enumerate}
 \item There is a function $\lambda: E \longrightarrow \mathbb{R}$ such that
 \begin{equation} \label{z}
 \forall j \in V,  \quad  x(j) = \sum_{j' \in V: (j', j) \in E}   \lambda_{(j', j)}  - \sum_{j' \in V: (j, j') \in E}   \lambda_{(j, j')}. 
 \end{equation}
 \item For each function $w: V \longrightarrow \mathbb{R}$,
$$\sum_{j \in V} x(j) w(j) = \sum_{ j' \in V: (j, j') \in E} \lambda(j, j') \bigl (w(j') - w(j) \bigr).$$
\end{enumerate}  
  \end{lemma}

{\it Proof}  

\noindent {\bf Point 1}. The proof proceeds by induction on the number of vertices of $\Gamma$. \\

{\sc Initialization}: If $V$ contains exactly one element, there is no directed edge and the assertions follow trivially. \\

{\sc Induction hypothesis}: Assume that the assertion holds for all acyclic directed graphs with only one component of size at most $p\geq 1$. \\

{\sc Induction step}: Consider an acyclic digraph $\Gamma = (V, E)$ with only one component of size $p +1$
and such that there is $x: V \longrightarrow \mathbb{R}$ as hypothesized.
 Note that there exists a vertex of $\Gamma$, say $j_{0}$, whose deletion ensures that the resulting acyclic digraph contains only one component and such that $j_0$ is the tail of no directed edge.
Let $d_{j_0}\geq 1$ be the number of directed edges whose $j_0$ is the head. Let $G_{-j_0} = (V \setminus {j_0}, E_{-j_0})$ be the resulting acyclic digraph without $j_0$ and its $d_{j_0}$ directed edges. Define the function $x': V \setminus \{i_0\} \longrightarrow \mathbb{R}$ as follows:

$$x'(j) = x(j) +   \frac{x(j_0)}{d_{j_0}} \quad \forall j \in V: (j, j_0) \in E, \mbox { and }   x'(j)  = x(j) \mbox{ otherwise}. $$  
Obviously,  
$$\sum_{j \in V \setminus j_0} x' (j) = \biggl ( \sum_{j \in V \setminus j_0} x(j) \biggr) +  x(j_0) =  \sum_{j \in V} x(j) = 0. $$
Thus,  by the induction hypothesis,  there is a function $\lambda: E_{-j_0} \longrightarrow \mathbb{R}$ such that
$$  \forall j \in V \setminus j_0, \quad x'(j) = \sum_{j' \in V \setminus j_0 : (j', j) \in E_{j_0}}   \lambda_{(j, j')} - \sum_{j' \in V \setminus j_0: (j, j') \in E_{j_0} }  \lambda_{(j, j')} .$$
One extends the function $\lambda$ from the domain $E_{-j_0}$ to the domain $E$ by setting
$$\forall j \in V : (j, j_0) \in E, \quad  \lambda(j, j_0) = \frac{x(j_0)}{d_{j_0}}.$$
Because there is no directed edge $(j_0, j)$ in $E$ by definition of $j_0$, one obtains
$$ x(j_0) = \sum_{j \in V: (j, j_0) \in E }\lambda(j, j_0), $$
and for each $j \in V$ such that $(j, j_0) \in E$,
\begin{eqnarray}
 x(j) &=& x'(j) -   \frac{x(j_0)}{d_{j_0}} \nonumber \\
&=&  \sum_{j' \in V \setminus j_0 : (j', j) \in E_{j_0}}   \lambda_{(j, j')} - \sum_{j' \in V \setminus j_0: (j, j') \in E_{j_0} }  \lambda_{(j, j')} -  \lambda(j, j_0)  \nonumber \\
&=&  \sum_{j' \in V : (j', j) \in E_{}}   \lambda_{(j, j')} - \sum_{j' \in V: (j, j') \in E_{} }  \lambda_{(j, j')}. \nonumber
\end{eqnarray} 
Obviously, for each other vertex $j$ equation (\ref{z}) holds. From this, conclude that the assertion of {\bf Point 1} is true.

\noindent {\bf Point 2}. 
From {\bf Point 1},
 $$ \sum_{ j \in V } x(j) w(j)  =\sum_{j \in V } \biggl( \sum_{j'\in V: (j', j) \in E} \lambda_{(j', j)} - \sum_{j' \in V: (j, j') \in E}   \lambda_{(j, j')}\biggr) w(j) .$$
 For each directed edge $(j, j')\in E$, the quantity $\lambda_{(j, j')}$ appears twice in the above equation, once for $j$ and once for $j'$, so that
  $$ \sum_{ j \in V } x(j) w(j)  = \sum_{(j, j') \in E} \lambda_{(j, j')} \biggl (w(j') - w(j)\biggr),$$
  as desired.
 \qed

 Before proving Theorem  \ref{th:marg_cov}, we need two additional lemmas.

 \begin{lemma}  \label{lem:cc}
 Consider any $i \in N$ and any component $C$ of
$\Gamma_{\cal F}^i$ not containing $\emptyset_M$. If a value  $\Phi$ on ${\cal G}_{N, {\cal P}, {\cal F}}$ satisfies the Null agent axiom, then it holds that:
$$  \Phi_i \biggl(N, \sum_{{\cal K} \in C } 1_{{\cal K}}, {\cal P}, {\cal F}  \biggr)   = 0.  $$
   \end{lemma}

{\it Proof} Pick any directed edge $({\cal K}', {\cal K}'' ) \in E_{\cal F}^i$.
By definition of a component, only two cases arise. Either $\{ {\cal K}', {\cal K}'' \} \subseteq C$ or $\{ {\cal K}', {\cal K}'' \} \cap C= \emptyset$. In the first case,
  $$\sum_{{\cal K} \in C } 1_{{\cal K}} ({ \cal K}') = \sum_{{\cal K} \in C }  1_{\cal K} ( {\cal K}'') = 1.$$
 In the second case,
    $$\sum_{{\cal K} \in C } 1_{{\cal K}} ({ \cal K}') = \sum_{{\cal K} \in C }  1_{\cal K} ( {\cal K}'') = 0.$$
   Therefore, $i$ is a null agent in this game, and so the result follows by the Null agent axiom.
\qed

\begin{lemma} \label{lem:S_emptyset}
 Consider any $i \in N$ with a component $C_{\emptyset_{M}}^{}$ of
$\Gamma_{\cal F}^i$ containing $\emptyset_M$. For each $q \in M$, define the subset of feasible coalitions $D^i_q \subseteq {\cal F}_q$ containing $i,$ and which covers the empty coalition, 
plus the empty coalition, that is,
 $$D^i_q  = \biggl \{K \in {\cal F}_q:  (\emptyset,  K) \in E_{{\cal F}_q}, i \in  K    \biggr\} \cup \bigl  \{\emptyset \bigr \}.$$ 
Define  the cartesian product $D^i$  of the sets $D^i_q$, $q \in M$. Then, 
 $C_{ \emptyset_{M} }^{} = D^i.$
\end{lemma}

{\it Proof}
We first prove that  $D^i \subseteq C_{ \emptyset_{M} }^{}$. Pick any
 ${\cal K} \in D^i.$ The proof proceeds by induction on the cardinality of the support $\mu ({\cal K})$. \\
 
 {\sc Initialization}: If $\mu ({\cal K})$ is empty, then ${\cal K} = \emptyset_{M} \in C_{ \emptyset_{M} }^{}$. \\

{\sc Induction hypothesis}: Assume that the assertion holds for $\mu ({\cal K})$ with at most $p\geq 0$ elements. \\

{\sc Induction step}: Consider the situation where  $\mu ({\cal K})$ contains $p + 1$ elements.  Pick any component $q \in M$ such that $K_q \not =\emptyset$ in ${\cal K}$. Consider the coalition profile
$(\emptyset, {\cal K}_{-q})$ whose support $\mu(\emptyset, {\cal K}_{-q})$ contains exactly $p$ elements. By the induction hypothesis $(\emptyset, {\cal K}_{-q}) \in C_{ \emptyset_{M} }^{}$. On the other hand,
by definition of ${\cal K}$,
$(\emptyset, K_q) \in E_{{\cal F}_q}$ and $i \in K_q$. Therefore,  ${\cal K}$ covers $(\emptyset, {\cal K}_{-q})$ with respect to $\subseteq^M$ and so  $((\emptyset, {\cal K}_{-q}),{\cal K}) \in E^{i}_{\cal F}$.
All in all, $(\emptyset, {\cal K}_{-q}) \in C_{ \emptyset_{M} }^{}$ and  $((\emptyset, {\cal K}_{-q}),{\cal K}) \in E^{i}_{\cal F}$ imply ${\cal K} \in C_{ \emptyset_{M} }^{}$, as desired.\\

Thus, the inclusion $D^i \subseteq C_{ \emptyset_{M} }^{}$ holds. Assume to the contrary that this inclusion is strict. Then,
there exists a direct edge $({\cal K}, {\cal K}')  \in  E^{i}_{\cal F}$
with one of the endpoints in $D^i$ and the other one out of $D^i$. But this is clearly impossible by definition of  $D^i$.  Thus, 
$D^i = C_{ \emptyset_{M} }^{}$.
 \qed

{\it  Proof }[(of Theorem  \ref{th:marg_cov})]
It is clear that any marginalist value as defined in (\ref{Marg}) satisfies Linearity and the Null agent axiom. So, 
let $\Phi$ be a value satisfying Linearity and the  Null agent axiom. To show:  $\Phi$ is a marginalist value as defined in (\ref{Marg}).  Pick any $(N, v,{\cal P}, {\cal F} ) \in {\cal G}_{N, {\cal P}, {\cal F}}$ and any   $i \in N$. 
By Linearity,
$$\Phi_i(N, v,{\cal P}, {\cal F}) = \sum_{{\cal K} \in {\cal F} \setminus \emptyset^M }v({\cal K}) \Phi_i(N, 1_{{\cal K}}, {\cal P}, {\cal F}) . $$
If ${\cal K} $ is not a vertex of the subdigraph $\Gamma_{\cal F}^i$, then $i$ is null in  $(N, 1_{{\cal K}}, {\cal P}, {\cal F})$ and so, by the Null agent  axiom,  $\Phi_i(N, 1_{{\cal K}}, {\cal P}, {\cal F}) = 0$. It follows that
\begin{equation} \label{eqgene}
\Phi_i(N, v,{\cal P}, {\cal F})  =  \sum_{ C \in {\cal C}_{\Gamma^i_{\cal F},} \atop  C \neq C_{\emptyset} } \sum_{{\cal K} \in C}  v({\cal K}) \Phi_i(N, 1_{{\cal K}}, {\cal P}, {\cal F})  +
  \sum_{{\cal K} \in C_{\emptyset}, \atop {\cal K} \neq \emptyset_M}  v({\cal K}) \Phi_i(N, 1_{{\cal K}}, {\cal P}, {\cal F}),
  \end{equation}
where  $C_{\emptyset}$ is the (possibly empty) component of ${\cal C}_{\Gamma^i_{\cal F}}$ containing the empty coalition profile $\emptyset_{M}$.
By Lemma \ref{lem:cc}, agent $i$ is null in the game 
$$(N,  \sum_{{\cal K} \in C_{}} 1_{{\cal K}}, {\cal P}, {\cal F}) \,\, \mbox{ and one obtains } \,\, \Phi_i(N,  \sum_{{\cal K} \in C_{}} 1_{{\cal K}},  {\cal P}, {\cal F}) = 0.  $$
By Linearity of $\Phi$,
$$\sum_{{\cal K} \in C_{}} \Phi_i(N, 1_{{\cal K}}, {\cal P}, {\cal F}) = 0. $$
Therefore, using Lemma \ref{lem:combi}, it holds that for each $C  \in {\cal C}_{\Gamma^i_{\cal F}}$,  $C \neq C_{{\emptyset}_M}$, 
$$ \sum_{{\cal K} \in C}  v({\cal K}) \Phi_i(N, 1_{{\cal K}}, {\cal P}, {\cal F}) =  \sum_{({\cal K}, {\cal K}') \in E^i_{{\cal F}} (C)} \lambda_{i}^{\Phi} ({\cal K}, {\cal K'}) \bigl(v( {\cal K}') - v ({\cal K}) \bigr).  $$
 for some function $\lambda_{i }^{\Phi} : E^i_{\cal F}(C) \longrightarrow \mathbb{R}$, where $E^i_{\cal F}(C)$ stands for the subset of directed edges in the component $C$ of $\Gamma^i_{{\cal F}} = (V^i_{\cal F}, E^i_{\cal F}) $.
Therefore, $\Phi_i(N, v,{\cal P}, {\cal F})$  in (\ref{eqgene}) can be rewritten as:

\begin{equation} \label{equgenef}
  \sum_{ C \in {\cal C}_{\Gamma^i_{\cal F}}, \atop  C \neq C_{{\emptyset}_M} }  \sum_{({\cal K}, {\cal K'}) \in E^i_{{\cal F}} (C)} \lambda_{i}^{\Phi} ({\cal K}, {\cal K'}) \biggl (v( {\cal K}') - v ({\cal K}) \biggr)   +
  \sum_{{\cal K} \in C_{\emptyset}, \atop {\cal K} \neq \emptyset_M}  v({\cal K}) \Phi_i(N, 1_{{\cal K}}, {\cal P}, {\cal F}).
  \end{equation} 
To finish the proof, consider any ${\cal K} \in C_{\emptyset} \setminus\emptyset_M$.
By Lemma \ref{lem:S_emptyset},  there exists at least one path  $({\cal K}_{0}^{\cal K} , {\cal K}_{1}^{\cal K}, \ldots, {\cal K}^{\cal K}_q)$ in the component $C_{\emptyset}$ of $\Gamma_{\cal F}^i$
from ${\cal K}_{0}^{\cal K} = \emptyset_M$ to ${\cal K}_{q}^{\cal K} = {\cal K}$. Since
$v(\emptyset_M) = 0$, it holds that: 
 $$v({\cal K}) \Phi_i(N, 1_{\cal K}, {\cal P}, {\cal F}) = \Phi_i(N, 1_{\cal K}, {\cal P}, {\cal F})   \sum_{r= 0}^{q-1}    \biggl(v({\cal K}_{r +1}^{\cal K})  -  v({\cal K}_{r}^{\cal K})\biggr) . $$
 Therefore, $\Phi_i(N, v,{\cal P}, {\cal F})$  in (\ref{equgenef}) can be rewritten as
 \begin{eqnarray}
 \sum_{ C \in {\cal C}_{\Gamma^i_{\cal F}}, \atop  C \neq C_{\emptyset} }  \sum_{({\cal K}, {\cal K'}) \in E^i_{{\cal F}} (C)} \lambda_{i} ({\cal K}, {\cal K'}) \biggl (v( {\cal K}') - v ({\cal K}) \biggr) & &   \nonumber \\
 +
  \sum_{{\cal K} \in C_{\emptyset}, \atop {\cal K} \neq \emptyset_M}  \Phi_i(N, 1_{\cal K}, {\cal P}, {\cal F})   \sum_{r= 0}^{q-1}    \biggl(v({\cal K}_{r +1}^{\cal K})  -  v({\cal K}_{r}^{\cal K})\biggr), \nonumber   & &  \nonumber \\
  \end{eqnarray}
 from which one concludes that $\Phi$ can be expressed as a marginalist value as in $(\ref{Marg})$ by extending the function $\lambda_{i}^{\Phi}$ to $E_{\cal F}^i(C_{\emptyset})$. 
Indeed, consider any directed edge 
 $({\cal K}, {\cal K}') \in  {E_{\cal F}}^i(C_{\emptyset})$ and let $X({\cal K}, {\cal K}')$  be the subset of coalition profiles ${\cal K}'' \in  C_{\emptyset} \setminus \emptyset_M$  whose selected path from $\emptyset_M $ to ${\cal K}''$ passes through the directed edge $({\cal K}, {\cal K}') $. From the previous equality, one obtains: 
\begin{eqnarray}
\Phi_i(N, v, {\cal P}, {\cal F}) &  =  & \sum_{ C \in {\cal C}_{\Gamma^i_{\cal F}}, \atop  C \neq C_{{\emptyset}_M} }  \sum_{({\cal K}, {\cal K'}) \in E^i_{{\cal F}} (C)} \lambda_{i} ({\cal K}, {\cal K'}) \bigl (v( {\cal K}') - v ({\cal K}) \bigr) \nonumber \\
&    &  +
  \sum_{({\cal K}, {\cal K}') \in  {E_{\cal F}}^i(C_{\emptyset})}  \biggl(  \sum_{{\cal K}''  \in X({\cal K}, {\cal K}')}  \Phi_i(N, 1_{\cal K''},  {\cal P}, {\cal F}) \biggr) \bigl(v({\cal K'})  -  v({\cal K}) \bigr),  \nonumber
  \end{eqnarray}
which ensures that $\Phi$  is marginalist. 
 \qed

By adding Efficiency to the statement of Theorem \ref{th:marg_cov} one obtains the family of flow methods on ${\cal G}_{N, {\cal P}, {\cal F}}$.

\begin{theorem} \label{flow-Eff} Consider any marginalist value $\Phi$ on ${\cal G}_{N, {\cal P}, {\cal F}}$. The value $\Phi$ is Efficient if and only if it is a flow method as defined in (\ref{Marg}) and (\ref{flowM}). 
\end{theorem}

{\it Proof} Assume first that $\Phi$ is an Efficient marginalist value. As  $\Phi$ is marginalist,  there are functions
$\lambda_{i}^{\Phi}: E_{\cal F}^i \longrightarrow \mathbb{R}$, 
$i \in N$, such that
\begin{equation} \label{eqMarg}
\forall v: {\cal F} \longrightarrow \mathbb{R}, \,\, \forall i \in N, \Phi_i(N, v, {\cal P}, {\cal F}) = \sum_{({\cal K}, {\cal K}') \in E_{\cal F}^i} \lambda_{i}^{\Phi}({\cal K}, {\cal K}') \bigl(v( {\cal K}') - v({\cal K}) \bigr).
\end{equation}
Consider the function $\Lambda_{}^{\Phi}:  E_{\cal F} \longrightarrow \mathbb{R}$ as in (\ref{flowM}).

In particular, for $v = 1_{\cal K}$,      ${\cal K} \in {\cal F} \setminus \{{\emptyset_{M}, {\cal P}}\}$, expression (\ref{eqMarg}) becomes
$$\Phi_i(N, v, 1_{\cal K}, {\cal F}) =  \sum_{{\cal K}' \in {\cal F}: \atop ({\cal K}, {\cal K}') \in E_{\cal F}^i} \lambda_{i}^{\Phi}({\cal K}, {\cal K}') (0 - 1) +  \sum_{{\cal K}' \in {\cal F}: \atop({\cal K}', {\cal K}) \in E_{\cal F}^i} \lambda_{i}^{\Phi}({\cal K}', {\cal K}) (1 - 0).$$
Taking the sum over all agents in $N$, one obtains
\begin{eqnarray} \label{eqMarg2}
\sum_{i \in N} \Phi_i(N, 1_{\cal K}, {\cal P},  {\cal F}) &=& - \sum_{i \in N} \sum_{{\cal K}' \in {\cal F}: \atop ({\cal K}, {\cal K}') \in E_{\cal F}^i} \lambda_{i} ^{\Phi}({\cal K}, {\cal K}')  +  \sum_{i \in N} \sum_{{\cal K}' \in {\cal F}: \atop ({\cal K}', {\cal K}) \in E_{\cal F}^i} \lambda_{i}^{\Phi}({\cal K}', {\cal K}) \nonumber \\
&= &\! -\!\!\!\! \sum_{{\cal K}' \in {\cal F}: \atop ({\cal K}, {\cal K}') \in E_{\cal F}} \sum_{i \in Q_{ ({\cal K}, {\cal K}') }}  \lambda_{i}^{\Phi}({\cal K}, {\cal K}') + \!\!\!\!  \sum_{{\cal K}' \in {\cal F}: \atop ({\cal K}', {\cal K}) \in E_{\cal F}} \!\!\!  \sum_{i \in Q_{ ({\cal K}', {\cal K}) }}  \lambda_{i}^{\Phi}({\cal K}', {\cal K}) \nonumber \\
 &=& -  \sum_{{\cal K}' \in {\cal F}: \atop ({\cal K}, {\cal K}') \in E_{\cal F}} \Lambda_{}^{\Phi} ({\cal K}, {\cal K}')  + \sum_{{\cal K}' \in {\cal F}: \atop ({\cal K}', {\cal K}) \in E_{\cal F}} \Lambda_{}^{\Phi} ({\cal K}', {\cal K}).
\end{eqnarray} 
By Efficiency,
$$\sum_{i \in N} \Phi_i(N, 1_{\cal K}, {\cal P}, {\cal F}) = v ({\cal P}) = 0,$$
so that (\ref{eqMarg2}) becomes
$$\sum_{{\cal K}' \in {\cal F}: \atop ({\cal K}, {\cal K}') \in E_{\cal F}} \Lambda_{}^{\Phi} ({\cal K}, {\cal K}') =  \sum_{{\cal K}' \in {\cal F}: \atop ({\cal K}', {\cal K}) \in E_{\cal F}} \Lambda_{}^{\Phi} ({\cal K}', {\cal K}),$$
which means that the conservation constraints are satisfied. This proves that the function  $\Lambda^{\Phi}$ is a flow on the product digraph $\Gamma_{\cal F} = ({\cal F}, E_{\cal F})$. To show that the latter is unitary,
consider the game $(N, v, 1_{\cal P}, {\cal F})$. Proceeding as above, noting there is no outgoing edge of ${\cal P}$, and applying Efficiency, one obtains
\begin{eqnarray}
 1 &=  &1_{\cal P} ({\cal P}) \nonumber \\
 &=&  \sum_{i \in N} \Phi_i(N, v, 1_{\cal P}, {\cal F}) \nonumber \\
 & = & \sum_{i \in N}  \sum_{({\cal K}, {\cal P}) \in E^i_{\cal F}} \lambda_{i}^{\Phi} ({\cal K}, {\cal P}) (1 - 0)  \nonumber \\
 & =&  \sum_{({\cal K}, {\cal P}) \in E_{\cal F}} \sum_{i \in Q_{({\cal K}, {\cal P})}} \lambda_{i}^{\Phi} ({\cal K}, {\cal P}) \nonumber \\
&  = &\sum_{{\cal K} \in {\cal F}:\atop ({\cal K}, {\cal P}) \in E_{\cal F}} \Lambda_{}^{\Phi} ({\cal K}, {\cal P}), \nonumber
 \end{eqnarray}
showing the result.

Conversely, assume that the marginalist value $\Phi$ is a flow method. Then, there is a function   $\Lambda_{}^{\Phi}:  E_{\cal F} \longrightarrow \mathbb{R}$ that induces a flow on the product digraph $\Gamma_{\cal F}$ as in (\ref{flowM}). To show:
$$\sum_{i \in N} \Phi_i(N, v, {\cal P}, {\cal F}) = v ({\cal P}).$$
Consider any coalition profile ${\cal K} \in {\cal F}$ different from ${\cal P}$ and $\emptyset_{M}$ and compute the coefficient $\gamma ({\cal K})$ of $v({\cal K})$ in the above total sum of payoffs:
\begin{eqnarray}
\gamma ({\cal K}) & =& \sum_{i \in N} \biggl ( \sum_{{\cal K}' \in {\cal F}: \atop ({\cal K}', {\cal K}) \in E_{\cal F}^i }  \lambda_{i}^{\Phi}({\cal K}', {\cal K}) -  \sum_{{\cal K}' \in {\cal F}: \atop ({\cal K}, {\cal K}') \in E_{\cal F}^i}  \lambda_{i}^{\Phi}({\cal K}, {\cal K}') \biggr) \nonumber \\
&= &  \sum_{{\cal K}' \in {\cal F}: \atop ({\cal K}, {\cal K}') \in E_{\cal F}} \sum_{i \in Q_{ ({\cal K}, {\cal K}') }}  \lambda_{i}^{\Phi}({\cal K}, {\cal K}') +  \sum_{{\cal K}' \in {\cal F}: \atop ({\cal K}', {\cal K}) \in E_{\cal F}} 
 \sum_{i \in Q_{ ({\cal K}', {\cal K}) }}  \lambda_{i}^{\Phi}({\cal K}', {\cal K}) \nonumber \\
 &=& -  \sum_{{\cal K}' \in {\cal F}: \atop ({\cal K}, {\cal K}') \in E_{\cal F}} \Lambda_{}^{\Phi} ({\cal K}, {\cal K}')  + \sum_{{\cal K}' \in {\cal F}: \atop ({\cal K}', {\cal K}) \in E_{\cal F}} \Lambda_{}^{\Phi} ({\cal K}', {\cal K}). \nonumber
\end{eqnarray}
As $\Lambda_{}^{\Phi}:  E_{\cal F} \longrightarrow \mathbb{R}$ defines a flow on $\Gamma_{\cal F}$,  it satisfies the conservation constraints so that $\gamma ({\cal K}) = 0$.
Proceeding in a similar way for $\gamma ({\cal P})$, one obtains
$$ \gamma ({\cal P})  =  \sum_{{\cal K} \in {\cal F}: \atop ({\cal K}, {\cal P}) \in E_{\cal F}} \Lambda_{}^{\Phi} ({\cal K}, {\cal P}).$$
Because the flow  $\Lambda^{\Phi}$ is unitary,   $\gamma ({\cal P}) = 1$. Taking into account the fact that 
$v(\emptyset_M) = 0$, one obtains
$$\sum_{i \in N} \Phi (N, v, {\cal P}, {\cal F}) = \sum_{{\cal K} \in {\cal F}} \gamma ({\cal K}) v({\cal K}) =    v({\cal P}),$$
showing that $\Phi$ is Efficient.
\qed
 

%
%

 \section{Two-step flow methods}  \label{Section-Two-step-FM}
 
 When agents are arranged in a coalition configuration, the payoff allocation procedure can be designed as a two-step procedure. In the first step, the procedure settles the payoff allocation between the elements of the coalition configuration. In the second step, the procedure settles the payoff allocation between the agents in each element of the coalition configuration, taking into account the result of the first step. 
In the particular case where the coalition configuration ${\cal P}$ is a partition of the agent set $N$ and the set function $v$ is defined over $2^N$, that is $(N, v)$ is a TU-game without restricted cooperation,
Owen \cite{Owen1977} proposes such a procedure by applying the Shapley value (Shapley \cite{Shapley1953}) at each step. The resulting value is known as the Owen value.  Note that the Shapley value is a flow method as it satisfies the principles of null agent, linearity and efficiency (see Weber \cite{Weber1988}).
Albizuri et al. \cite{Albizuri2006a} apply Owen's procedure to situations where ${\cal P}$ is not necessarily a partition of $N$, but maintain the assumption that $(N, v)$ is a TU-game without restricted cooperation. The resulting value is the configuration value. 
Two-step procedures are subsequently used in a broader context where the TU-game $(N, v)$ is augmented by a partition ${\cal P}$ and a communication graph on $N$ that represents the bilateral communication constraints among the agents. In this context, different values can be applied at each step such as the Myerson value (Myerson \cite{Myerson1977}) or the Average tree solution (Herings et al. \cite{Herings2008}) (see, e.g., V\'azquez-Brague et al. \cite{Vazquez1996}, Brink van den et al. \cite{Brink-K-Laan}, Alonso-Meijide et al., \cite{Alonso-2009}, B\'eal et al. \cite{Beal2022}).

In this section, we design a two-step flow procedure for games in ${\cal G}_{N, {\cal P}, {\cal F}}$ where at each step of the procedure a flow method is applied. Pick any game $(N, v, {\cal P}, {\cal F}) \in {\cal G}_{N, {\cal P}, {\cal F}}$. Denote by $M = \{1, \ldots, m\}$ the index set of the elements of ${\cal P}$.
The procedure results in a two-step flow method where the flow passing through directed edges of the product digraph $\Gamma_{\cal F}$ can be decomposed into two types of flows: a flow on the directed edges of the digraph induced by the set system $(M, (\{q\})_{q \in M})$ and the flows passing trough the directed edges of each digraph $\Gamma_{{\cal F}_q}$ associated with the set system $(P_q, {\cal F}_q)$, $q \in M$. 

To construct two-step flow methods, we need to consider two other types of subdomains of ${\cal G}_{N, {\cal P}, {\cal F}}$. These domains are constructed from $N$, ${\cal P}$ and ${\cal F}$. To introduce them, some definitions are in order. Denote by ${\cal P} _M$ the finest partition  $(\{q\})_{q \in M}$ of $M$ and by ${\cal F}_M$ the cartesian product of the trivial set systems $(\emptyset, \{q\})$, $q \in M$. We introduce:

-- the subdomain ${\cal G}_{M,  {\cal P} _M, {\cal F}_M}$ of games of the form $(M, v_M,  {\cal P} _M, {\cal F}_M  )$, where only the coalition profile function $v_M: {\cal F}_M \longrightarrow \mathbb{R}$ can vary. Note that the corresponding digraph $\Gamma_{{\cal F}_M}$ is a directed hypercube of dimension $m$;

-- for each $q \in M$, the subdomain ${\cal G}_{P_q,  (P_q), {\cal F}_q}$ of games of the form $(P_q,  v_q, (P_q), {\cal F}_q)$ where only the coalition profile function $v_q: {\cal F}_q \longrightarrow \mathbb{R} $ can vary. \\
 
Following the usual terminology, games in ${\cal G}_{M,  {\cal P} _M, {\cal F}_M}$ are {\bf upper-games} of the original game and games in  ${\cal G}_{P_q,  (P_q), {\cal F}_q}$ are  {\bf $q$-lower games} of the original game. \\

\noindent {\bf Notation} Since on each of these subdomains the corresponding digraph is fixed once and for all, coefficient functions of a marginalist value $\Phi$ will be denoted by $\lambda_i^{\Phi}$ without mentioning the label of the digraph. The context should clearly indicate if $\Phi$, and so $\lambda_i^{\Phi}$, is a value on ${\cal G}_{N, {\cal P}, {\cal F}}$, ${\cal G}_{M,  {\cal P} _M, {\cal F}_M}$ or ${\cal G}_{P_q,  (P_q), {\cal F}_q}$.
\begin{flushright}
$\blacksquare$
\end{flushright}

\begin{example}
Consider again Example \ref{graph1}. The directed hypercube $\Gamma_{{\cal F}_M}$ is as follows: $M = \{1, 2\}$ and the vertices of $\Gamma_{{\cal F}_M}$  are  
$(\emptyset, \emptyset)$, $(\{1\}, \emptyset)$, $(\emptyset, \{2\})$, and $(\{1\}, \{2\})$; so it is the directed hypercube of dimension 2 represented in Figure \ref{fig-Hypercube}.

 \begin{figure}[h]
\begin{center}
\begin{tikzpicture} [every node/.style={scale = 0.6}, scale = 0.6]
 \node (A) at (0,0) {$(\emptyset, \emptyset)$};
  \node (B) at (5,0) {$(\emptyset, \{2\})$};
  \node (C) at (2.5, 1.5) {$(\{1\}, \emptyset)$};
  \node (D) at (7.5, 1.5) {$(\{1\}, \{2\})$};

  \draw[->] (A) -- (B);
  \draw[->] (B) -- (D);
  \draw [->] (C) -- (D);
  \draw [->] (A) -- (C);

\end{tikzpicture}
\end{center}
\caption{The directed hypercube $\Gamma_{{\cal F}_{M}}$ for $M = \{1, 2\}$.} 
\label{fig-Hypercube}
\end{figure}
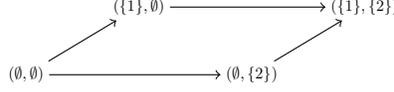
 \end{example}

Next, consider the coalition profiles  ${\cal K}=(K_1, \ldots, K_m) \in {\cal F}$ of support $\mu({\cal K}) = S \subseteq M$. Among this set of coalition profiles, the coalition profiles  ${\cal K}_{S, K_q}$ denotes the coalition profiles where $q \in  S$, $K_r = P_r$ for $r \in S \setminus \{q\}$,  $K_q \in {\cal F}_{q}$, and  $K_r = \emptyset$ otherwise.

\begin{remark} \label{notation}
 For the sake of simplicity, we use the notation ${\cal K}_{S, K_q}$ for $K_q = \emptyset$ and $q \in S$, 
 to mean that the coordinate $q$ is active even if it is equal to the empty coalition. So, we still consider that the support of ${\cal K}_{S, K_q}$ is $S$.
 \end{remark}
In the special case where $S = \{q\}$, ${\cal K}_{S, K_q} = (K_q, (\emptyset_M)_{-q})$. 
In case, $K_q = P_q$, simply denote this coalition profile by ${\cal K}_S$. In particular,  ${\cal K}_{\emptyset} = \emptyset_M$ and  ${\cal K}_{M} = {\cal P}$.
Coalition profiles of the form ${\cal K}_{S, K_q}$ for $K_q \in {\cal F}_q$, $S \subseteq M$ and $S \ni q$ are named {\bf relevant profiles}.
The set of {\bf relevant directed edges} of the product digraph $\Gamma_{\cal F} = ({\cal F}, E_{\cal F})$ is the subset of directed edges $E_{\cal F}^R \subseteq E_{\cal F}$ whose endpoints are relevant.

\begin{example}
Consider again Example \ref{graph1} and the product digraph  $\Gamma_{{\cal F}} = ({\cal F}, E_{{\cal F}})$. Figure \ref{relevant-graph2} represents the subdigraph induced by the set of relevant directed edges  $E_{\cal F}^R \subseteq E_{\cal F}$, and the subdigraph of $\Gamma_{{\cal F}}^3 = (V^3, E^3_{\cal F})$ induced by the subset of relevant directed edges  $E_{\cal F}^R \cap E^3_{\cal F}$.
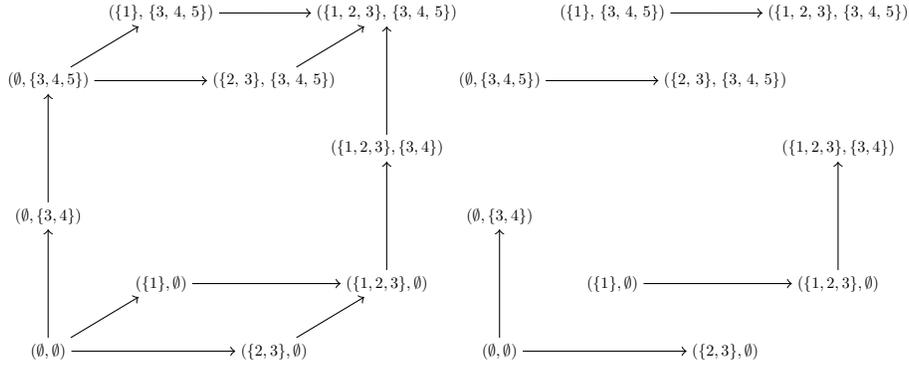
\begin{figure}[h]

\begin{center}
\begin{tikzpicture} [every node/.style={scale = 0.60}, scale = 0.60]

  \node (A) at (0,0) {$(\emptyset, \emptyset)$};
  \node (B) at (5,0) {$(\{2, 3\}, \emptyset)$};
  \node (C) at (2.5, 1.5) {$(\{1\}, \emptyset)$};
  \node (D) at (7.5, 1.5) {$(\{1, 2, 3\}, \emptyset)$};

  \draw[->] (A) -- (B);
  \draw[->] (B) -- (D);
  \draw [->] (C) -- (D);
  \draw [->] (A) -- (C);

 \node (E) at (0, 3) {$(\emptyset, \{3, 4\})$};
\node (H) at (7.5, 4.5) {$(\small \{1, 2, 3\}, \{3, 4\})$};  
  
  
  \node (I) at (0, 6) {$(\emptyset, \{3, 4, 5\})$};
\node (J) at (5, 6) {(\{2, 3\}, \{3, 4, 5\})};
\node (K) at (2.5, 7.5) {(\{1\}, \{3, 4, 5\})};
\node (L) at (7.5, 7.5) {(\{1, 2, 3\}, \{3, 4, 5\})};

   \draw [->] (I) -- (J);
 \draw [->] (I) -- (K);
 \draw [->] (J) -- (L);
 \draw [->] (K) -- (L);  

\draw [->] (A) -- (E);
\draw [->] (E) -- (I);

  
  
   \draw [->] (D) -- (H);
  \draw [->] (H) -- (L);  
  
  
    \node (A) at (10,0) {$(\emptyset, \emptyset)$};
  \node (B) at (15,0) {$(\{2, 3\}, \emptyset)$};
  \node (C) at (12.5, 1.5) {$(\{1\}, \emptyset)$};
  \node (D) at (17.5, 1.5) {$(\{1, 2, 3\}, \emptyset)$};

  \draw[->] (A) -- (B);
  \draw [->] (C) -- (D);

 \node (E) at (10, 3) {$(\emptyset, \{3, 4\})$};
\node (H) at (17.5, 4.5) {$(\small \{1, 2, 3\}, \{3, 4\})$};  
  
  
  \node (I) at (10, 6) {$(\emptyset, \{3, 4, 5\})$};
\node (J) at (15, 6) {(\{2, 3\}, \{3, 4, 5\})};
\node (K) at (12.5, 7.5) {(\{1\}, \{3, 4, 5\})};
\node (L) at (17.5, 7.5) {(\{1, 2, 3\}, \{3, 4, 5\})};

   \draw [->] (I) -- (J);
 \draw [->] (K) -- (L);  

\draw [->] (A) -- (E);

  
  
   \draw [->] (D) -- (H);
 
  \end{tikzpicture}
\end{center}
\caption{To the left, the subdigraph of $\Gamma_{{\cal F}} = ({\cal F}, E_{{\cal F}})$ induced by the set of relevant directed edges $E^R_{{\cal F}}$. To the right, the subdigraph of $\Gamma_{\cal F }^3 = (V^3_{\cal F}, E_{\cal F}^{3})$ induced by  relevant directed edges $E^R_{{\cal F}} \cap E_{\cal F}^{3}$. }
\label{relevant-graph2}
\end{figure}
\end{example}

\noindent {\bf Notation} To avoid any confusion between the coalition profiles in ${\cal F}$ and the coalition profiles in ${\cal F}_M$, we will henceforth use the notation ${\cal R}$ to denote a coalition profile in ${\cal F}_M$ instead of ${\cal K}$. To insist on the support of a coalition profile, we will often use the notation ${\cal R}_S$ to denote the coalition profile in ${\cal F}_M$ with support $S\subseteq M$.  
For instance, if $S = \{2, 3, 5\}$, then ${\cal R}_{\{2, 3, 5\}} = (\emptyset, \{2\}, \{3\}, \emptyset, \{5\}, \emptyset, \ldots, \emptyset)$. The coalition profile ${\cal R}_{\{2, 3, 5\}} $ will often be associated with a relevant coalition profile in ${\cal F}$ in the following manner: if ${\cal K} \in {\cal F}$ is a relevant coalition profile with support $\mu ({\cal K}) = S$, then the corresponding coalition profile in ${\cal F}_M$ is ${\cal R}_{S}$, that is, $({\cal R}_{S})_q = \{q\}$ if $q \in S$ and $({\cal R}_{S})_q = \emptyset$ otherwise. Thus, if
${\cal K} = (\emptyset, P_2, P_3, \emptyset, K_5, \emptyset, \ldots, \emptyset)$, then
${\cal R}_{\{2, 3, 5\}}$ indicates which elements of the coalition configuration form, without specifying that $K_5$ represents $P_5$.
\begin{flushright}
$\blacksquare$
\end{flushright}

Consider any game $(N, v, {\cal P}, {\cal F}) \in {\cal G}_{N,  {\cal P}, {\cal F}}$.
For each $q \in M$ and each feasible coalition $K_q  \in   {\cal F}_{q}$, define the {\bf $K_q$-upper game} $(M, v_{M}^{K_q}, {\cal P}_M,  {\cal F}_M ) \in {\cal G}_{M,  {\cal P} _M, {\cal F}_M}$ where 
$ v_{M}^{K_q}:  {\cal F}_M \longrightarrow \mathbb{R}$  is such that
$$\forall {\cal R}_S \in {\cal F}_M, \quad
 v_{M}^{K_q} ({\cal R}_S) =  \biggl \{ 
 \begin{array}{ll}
 v ({\cal K}_S)  & \mbox{if } S \subseteq M \setminus q,  \\
 v ({\cal K}_{S, K_q})&  \mbox {otherwise.}
  \end{array}
$$
This situation depicts a cooperative game played by the elements $P_r$, $r \in M$, of $ {\cal P}$ where $P_q$ is represented by the feasible coalition $K_q \in {\cal F}_{q}$.
When $K_q = P_q$, $v_{M}^{}$ stands for $v_{M}^{M}$.

\begin{remark} \label{reNull}
If $K_q = \emptyset$, then $q \in M$ is a null agent in the $K_q$-upper game  since, by definition, ${\cal K}_{S}$ coincides with ${\cal K}_{S, \emptyset}$. 
\end{remark}

 Denote by $\Phi^M$ a flow method on ${\cal G}_{M,  {\cal P} _M, {\cal F}_M}$ and by  $\Phi^q$ a flow method  on
 ${\cal G}_{P_q,  (P_q), {\cal F}_q}$, for $q \in M$. Next, we define a two-step flow method on
$ {\cal G}_{N,  {\cal P}, {\cal F}}$.
 
 A value $\Phi^{0}$ is a {\bf two-step flow method} on $ {\cal G}_{N,  {\cal P}, {\cal F}}$ if it exists 
 a  flow method $\Phi^M$ on ${\cal G}_{M,  {\cal P} _M, {\cal F}_M}$ and a flow method $\Phi^q$ on \
$ {\cal G}_{P_q,  (P_q), {\cal F}_q}$ for each $q \in M$ such that:
\begin{equation} \label{2-step-flow}
\forall i \in N, \quad  \Phi_i^{0} (N, v, {\cal P}, {\cal F}) = \sum_{q \in M: i \in P_q} \Phi_i^{q} (P_q, v_q, (P_q), {\cal F}_q),
\end{equation}
where
\begin{equation}  \label{2-step-flow1}
\forall q \in M, \forall K_q \in {\cal F}_q, \quad  v_q (K_q) =  \Phi^M_q (M, v_{M}^{K_q}, {\cal P}_M,  {\cal F}_M ).
\end{equation}

\begin{remark}  \label{coef}
In any upper-game, the digraph $\Gamma_{{\cal F}_M}$ corresponding to the set system
$(M, {\cal F}_M)$ is the $m$-dimensional directed hypercube. It follows that
the coefficient function $\lambda_q^{\Phi^M}: E_{ \Gamma_{{\cal F}_M} }^q\longrightarrow \mathbb{R}$ associated with the flow method $\Phi^{M}$ is such that
$$\forall {\cal R}_S \in {\cal F}_M: S \ni q, \quad \lambda_q^{\Phi^M} ({\cal R}_{S\setminus q}, {\cal R}_S) = \Lambda^{\Phi^M} ({\cal R}_{S\setminus q}, {\cal R}_S).$$
\end{remark}

\begin{remark}
 The coalition profile function $v_q : {\cal F}_M \longrightarrow \mathbb{R}$ in (\ref{2-step-flow1})  is well-defined since 
 $$v_q (\emptyset_M)   =  \Phi^M_q (M, v_{M}^{\emptyset}, {\cal P}_M,  {\cal F}_M ) = 0,$$
thanks to Remark \ref{reNull} and the fact that any flow method satisfies the null agent axiom. 
\end{remark}

So an agent's payoff in a game $(N, v, {\cal P}, {\cal F})$ is the sum of its payoffs in each $q$-lower game to which it belongs. For each $q$-lower game, the worth of a feasible coalition $K_q$ is defined by the payoff obtained by the element $P_q$ of the partition ${\cal P}$ is the $K_q$-upper game, where $K_q$ represents $P_q$. Furthermore, these payoffs result from flows passing through the directed edges of the digraphs $\Gamma_{{\cal F}_q}$ and the directed hypercube associated with $(M, {\cal F}_M)$.  

To show that the above procedure induces well-defined flow methods on ${\cal G}_{N,  {\cal P}, {\cal F}}$, we proceed in two steps. In a first step, we show that a value $\Phi^{0}$ defined through the two-step procedure (\ref{2-step-flow})-(\ref{2-step-flow1}) is a marginalist value. In a second step, we show that $\Phi^{0}$ is indeed a flow method. \\

\begin{lemma} \label{Marg-two-step} Let $\Phi^{0}$ be a value on  ${\cal G}_{N,  {\cal P}, {\cal F}}$  defined through the two-step procedure (\ref{2-step-flow})-(\ref{2-step-flow1}). Then, $\Phi^{0}$ is a marginalist value with coefficient functions
$\lambda_i^{\Phi^{0}}: E_{\cal F}^i \longrightarrow \mathbb{R}$, $i \in N$, defined as follows: for each  $({\cal K}, {\cal K}') \in E_{\cal F}^i,$
\begin{enumerate}
\item if $({\cal K}, {\cal K}') \not \in E_{\cal F}^{R} \cap E_{\cal F}^i$, then $\lambda_i^{\Phi^{0}} ({\cal K}, {\cal K}') =0$; 
\item if  $({\cal K}, {\cal K}') = ({\cal K}_{\mu ({\cal K}'), K_q}, {\cal K}'_{\mu ({\cal K}'), K'_q}) \in E_{\cal F} ^{R} \cap E_{\cal F}^i$,  $q \in \mu ({\cal K}')$, then 
$$\lambda_i^{\Phi^{0}} ({\cal K}, {\cal K}') = \Lambda^{\Phi^M} ({\cal R}_{\mu ({\cal K}') \setminus q}, {\cal R}_{\mu ({\cal K}')} ) \lambda_i^{\Phi^q} (K_q, K'_q).$$
\end{enumerate}
\end{lemma}

{ \it Proof }  Assume that $\Phi^{0}$ is a value defined through the two-step procedure (\ref{2-step-flow})-(\ref{2-step-flow1}). Pick any game $(N, v, {\cal P}, {\cal F}) \in {\cal G}_{N,  {\cal P}, {\cal F}}$ and
any $i \in N$. By definition  (\ref{2-step-flow})-(\ref{2-step-flow1}) of $\Phi^{0}$, for each element $P_q \ni i$ of ${\cal P}$, we have
\begin{eqnarray}  \label{eq-MargFlow}
 \Phi_i^{q} (P_q, v_q, (P_q), {\cal F}_q) &= &    \sum_{(K_q, K'_q) \in E^i_{{\cal F}_q}}  \lambda^q_i(K_q, K'_q)\bigl (v_q(K'_q) -v_q(K_q)\bigr).  \nonumber \\
  \end{eqnarray}
 Using Remark \ref{coef}, the right-hand side of  (\ref{eq-MargFlow})  writes
  \begin{eqnarray}   \label{eq-enplus}                                                             
   & & \sum_{(K_q, K'_q) \in E^i_{{\cal F}_q}}  \lambda^q_i(K_q, K'_q) \biggl (\Phi^M_q (M, v_{M}^{K_q'}, {\cal P}_M, {\cal F}_M) -  \Phi^M_q (M, v_{M}^{K_q}, {\cal P}_M, {\cal F}_M) \biggr)  \nonumber \\
                                                                                                     &=&       \sum_{(K_q, K'_q) \in E^i_{{\cal F}_q}}  \lambda^q_i(K_q, K'_q)  \biggl[  \sum_{S \subseteq M: S \ni q}   \Lambda^{\Phi^M}({\cal R}_{S\setminus q}, {\cal R}_{S})\bigl( v_{M}^{K'_q} ({\cal R}_S) -v_{M}^{K_q} ({\cal R}_S)\bigr) \biggr]
 \nonumber \\
 &=&       \sum_{(K_q, K'_q) \in E^i_{{\cal F}_q}}  \lambda^q_i(K_q, K'_q)  \biggl[  \sum_{S \subseteq M: S \ni q}     \Lambda^{\Phi^M}({\cal R}_{S\setminus q}, {\cal R}_{S}) \bigl( v({\cal K}_{S, K'_q}) -v({\cal K}_{S, K_q})\bigr) \biggr].  \nonumber \\
 \end{eqnarray}
 
Taking into account Remark \ref{notation}, for each pair  $((K_q, K'_q), {\cal R}_S) \in E^i_{{\cal F}_q} \times {\cal F}_M$ where $S \ni q$, there is a unique relevant directed edge $({\cal K}, {\cal K}') \in E^R \cap E^i_{\cal F} $ such that 
$\mu ({\cal K}') = S$, $K_q$ is the possibly empty coalition at coordinate $q$ of ${\cal K}$ and $K_q'$ is the nonempty coalition at coordinate $q$ of ${\cal K}'$. Conversely, for each relevant directed edge $({\cal K}, {\cal K}') \in E_{\cal F}^R \cap E^i_{\cal F} $ such that $q_{({\cal K}, {\cal K'})} = q$, there is a pair  $((K_q, K'_q), {\cal R}_S) \in E^i_{{\cal F}_q} \times {\cal F}_M$ such that $S \ni q$, $ \mu ({\cal K}') = S$, $K_q$ is possibly the empty coalition at coordinate $q$ of ${\cal K}$ and $K_q'$ is the coalition at coordinate $q$ of ${\cal K}'$. 
 Therefore, using (\ref{eq-enplus}), equation (\ref{eq-MargFlow}) can be rewritten as:
$$ \Phi_i^{q} (P_q, v_q, (P_q), {\cal F}_q) = \!\!\!\!\!\!\!\!\! \sum_{({\cal K}, { \cal K'}) \in E^R_{\cal F} \cap E_{{\cal F}}^i: \atop q_{({\cal K}, { \cal K'})} = q}  \lambda^q_i(K_q, K'_q)   \Lambda^{\Phi^M}({\cal R}_{\mu ({\cal K}') \setminus q}, {\cal R}_{\mu ({\cal K}')}) \bigl (v({\cal K}') - v({\cal K})\bigr).  $$
By definition,
\begin{eqnarray}
 \Phi^{0}_i (N, v, {\cal P}, {\cal F})  &= & \sum_ {q \in M : i \in P_q } \Phi_i^{q} (P_q, v_q, (P_q), {\cal F}_q), \nonumber
  \end{eqnarray} 
 so that  $\Phi^{0}_i (N, v, {\cal P}, {\cal F})$ is equal to
  $$\sum_ {q \in M : i \in P_q }  \sum_{({\cal K}, { \cal K'}) \in E^R_{\cal F} \cap E^i_{{\cal F}}: \atop q_{({\cal K}, { \cal K'})} = q}  \lambda^q_i(K_q, K'_q)   
\Lambda^{\Phi^M}({\cal R}_{ \mu ({\cal K}') \setminus q},{\cal R}_{ \mu ({\cal K}')} )\bigl (v({\cal K}') - v({\cal K})\bigr), $$

 which shows that $\Phi^{0}$ can be written as a marginalist value with coefficient functions $\lambda_i^{\Phi^{0}}: E_{\cal F}^i \longrightarrow \mathbb{R}$, $i \in N$, as in Point 1 and Point 2 of the statement of Lemma \ref{Marg-two-step}. 
\qed
From $\Phi^{0}$,  define the function $\Lambda^{\Phi^{0} }: E_{\cal F} \to \mathbb{R}$ as
\begin{equation} \label{flow-2step}
\forall ({\cal K}, { \cal K'}) \in E_{\cal F}, \quad  \Lambda^{\Phi^{0}} ({\cal K}, { \cal K'}) = \sum_{i \in Q_{( {\cal K}, { \cal K'})}} \lambda_i^{\Phi^{0}} ({\cal K}, { \cal K'}),
\end{equation} 
where $ \lambda_i^{\Phi^{0}} ({\cal K}, { \cal K'})$ is as in Point 1 and Point 2 of the statement of Lemma \ref{Marg-two-step}.
The following theorem shows that $\Lambda^{\Phi^{0} }$ is a unitary flow on $\Gamma_{\cal F}= ({\cal F}, E_{\cal F})$, which admits a decomposition into two different types of unitary flows: the first type of flow consists on the unitary flow on the $m$-dimensional directed hypercube induced by the flow method $\Phi^M$ applied to the upper games; the second type of flow corresponds to the unitary flows passing through the directed edges of the digraphs $\Gamma_{{\cal F}_q}$, $q \in M$. The latter flows are induced by the flow methods $\Phi^q$, $q \in M$, applied to the lower-games. Furthermore, if a directed edge of 
$\Gamma_{\cal F}$ is not relevant, then the flow passing through it is null.

\begin{theorem} \label{char-flowmethod}
 Let $\Phi^{0}$ be a value in ${\cal G}_{N,  {\cal P}, {\cal F}}$ defined through the two-step procedure (\ref{2-step-flow})-(\ref{2-step-flow1}). Then, $\Phi^{0}$ is a  flow method such that
 the flow $\Lambda^{\Phi^{0}}: E_{\cal F} \longrightarrow \mathbb{R}$ is defined as:
 \begin{enumerate}
 \item  $\forall ({\cal K}, {\cal K}') = ({\cal K}_{\mu ({\cal K}'), K_q}, {\cal K}_{\mu ({\cal K}'), K'_q}) \in E_{\cal F}^{R}, \, q \in \mu ({\cal K}')$,
 $$\Lambda^{\Phi^{0}} ({\cal K}, {\cal K}') = \Lambda^{\Phi^M} ({\cal R}_{\mu ({\cal K}') \setminus q}, {\cal R}_{\mu ({\cal K}')}) \Lambda^{\Phi^q} (K_q, K'_q); $$
 \item  $ \Lambda^{\Phi^{0}} ({\cal K}, {\cal K}') = 0$, otherwise.
 \end{enumerate}
 \end{theorem} 

{\it Proof}  Consider any value $\Phi^{0}$ in ${\cal G}_{N,  {\cal P}, {\cal F}}$  defined through the two-step procedure (\ref{2-step-flow})-(\ref{2-step-flow1}). By Lemma \ref{Marg-two-step},  $\Phi^{0}$ is a marginalist 
value with coefficient functions $\lambda_i^{0}$, $i \in N$, as in Point 1 and Point 2 of Lemma \ref{Marg-two-step}. It remains to prove that the function $\Lambda^{\Phi^{0}}: E_{\cal F} \longrightarrow \mathbb{R}$ defined as in (\ref{flow-2step}) is a unitary flow.  There are several cases to consider. \\

\noindent {\bf Case (a)}: if the coalition profile  ${\cal K} \in {\cal F}$ is not relevant. Then, by Point 1 of Lemma \ref{Marg-two-step},
$$\forall ({\cal K}, {\cal K}') \in E_{\cal F}, \quad  \Lambda^{\Phi^{0}} ({\cal K}, {\cal K}') = \sum_{i \in Q_{( {\cal K}, { \cal K'})}} \lambda_i^{\Phi^{0}} ({\cal K}, { \cal K}') =  \sum_{i \in Q_{( {\cal K}, { \cal K}')}}  0 = 0,$$
and
$$\forall ({\cal K}', {\cal K}) \in E_{\cal F}, \quad  \Lambda^{\Phi^{0}} ({\cal K}', {\cal K}) = \sum_{i \in Q_{( {\cal K'}, { \cal K})}} \lambda_i^{\Phi^{0}} ({\cal K}', { \cal K}) =  \sum_{i \in Q_{( {\cal K}, { \cal K'})}}  0 = 0,$$
so that the conservation constraints are trivially satisfied for any such coalition profile/vertex of  $\Gamma_{\cal F}$.

\noindent {\bf Case (b)}:  if the coalition profile  ${\cal K} \in {\cal F}$ is relevant, different from $\emptyset_M$, ${\cal P}$, and such for each $ q\in \mu ({\cal K})$, $K_q = P_q$, that is
${\cal K} = {\cal K}_{\mu ({\cal K})}$.  In this case, if $({\cal K}, {\cal K}') \in E_{\cal F}$, then it is a relevant directed edge and  $\mu ({\cal K}') = \mu ({\cal K}) \cup q$ for some $q \in M \setminus  \mu ({\cal K})$. From this observation, 
Lemma \ref{Marg-two-step} and the definition of the flows $ \Lambda^{\Phi^q}: E_{{\cal F}_q} \longrightarrow \mathbb{R}$, one obtains
\begin{eqnarray} \label{conserv-rel}
\sum_{({\cal K}, {\cal K}') \in E_{\cal F} } \Lambda^{\Phi^{0}} ({\cal K}, {\cal K}')& =&\sum_{({\cal K}, {\cal K}') \in E_{\cal F} } \sum_{i \in Q_{({\cal K}, {\cal K}')}} \lambda_i^{\Phi^0} ({\cal K}, {\cal K}') \nonumber \\
& = & \!\!\!\!\! \sum_{q \in M \setminus \mu ({\cal K})}  \Lambda^{\Phi^{M}} ({\cal R}_{\mu ({\cal K})}, {\cal R}_{\mu ({\cal K}) \cup q} ) \sum_{(\emptyset, K'_q) \in E_{{\cal F}_q}} \sum_{i \in K'_q}   \lambda_i^{\Phi^{q}} (\emptyset, K'_q)  \nonumber \\
&= &\!\!\!\!\! \sum_{q \in M \setminus \mu ({\cal K})}  \Lambda^{\Phi^{M}} ({\cal R}_{\mu ({\cal K})}, {\cal R}_{\mu ({\cal K}) \cup q} ) \sum_{(\emptyset, K'_q)\in E_{{\cal F}_q}} \Lambda^{\Phi^q} (\emptyset, K_q').  
\end{eqnarray} 
As   $\Lambda^{\Phi^q}$ defines a unitary flow on the digraph $\Gamma_{{\cal F}_q}$, (\ref{conserv-rel}) rewrites as
$$\sum_{({\cal K}, {\cal K}') \in E_{\cal F} } \Lambda^{\Phi^{0}} ({\cal K}, {\cal K}') =  \sum_{q \in M \setminus \mu ({\cal K})}  \Lambda^{\Phi^{M}} ({\cal R}_{\mu ({\cal K})}, {\cal R}_{\mu ({\cal K}) \cup q} ).$$ 
Proceeding in the same way for directed edges  $({\cal K}', {\cal K}) \in E_{\cal F}$, one obtains,
$$\sum_{({\cal K}', {\cal K}) \in E_{\cal F} } \Lambda^{\Phi^{0}} ({\cal K}', {\cal K}) =  \sum_{q \in \mu ({\cal K})} \Lambda^{\Phi^{M}} (\mu ({\cal R}_{\mu ({\cal K}) \setminus q}, {\cal R}_{\mu ({\cal K})} ).$$ 
As  $\Lambda^{\Phi^M}$ defines a unitary flow on the directed hypercube $\Gamma_{{\cal F}^M}$, it satisfies the conservation constraints and so
\begin{eqnarray}
\sum_{({\cal K}', {\cal K}) \in E_{\cal F} } \Lambda^{\Phi^{0}} ({\cal K}', {\cal K}) &= & \sum_{q \in \mu ({\cal K})} \Lambda^{\Phi^{M}} (\mu ({\cal R}_{\mu ({\cal K}) \setminus q}, {\cal R}_{\mu ({\cal K})} ) \nonumber \\
& = & \sum_{q \in M \setminus \mu ({\cal K})}  \Lambda^{\Phi^{M}} ({\cal R}_{\mu ({\cal K})}, {\cal R}_{\mu ({\cal K}) \cup q} ) \nonumber \\
&=& \sum_{({\cal K}, {\cal K}') \in E_{\cal F} } \Lambda^{\Phi^{0}} ({\cal K}, {\cal K}'). \nonumber
\end{eqnarray}
Thus, the flow induced by $\Lambda^{\Phi^{0}}$ and passing through such a ${\cal K}$ satisfies the conservation constraints.

\noindent {\bf Case (c)}:  Assume that the coalition profile  ${\cal K} \in {\cal F}$ is relevant, different from $\emptyset_M$, ${\cal P}$, and such that there is (a unique) $q \in \mu ({\cal K})$ such that $K_q \not =  P_q$, that is
${\cal K} = {\cal K}_{(\mu ({\cal K}), K_q)}$ for some $K_q \in {\cal F}_q \setminus \{\emptyset, P_q\}$ and $q \in \mu ({\cal K})$. Then, a  directed edge $({\cal K}, {\cal K}') \in E_{\cal F}$ is relevant if and only if
$ \mu ({\cal K}) =  \mu ({\cal K}')$, ${\cal K}'_{ - q } = {\cal K}_{ -q}$,  $(K_q, K'_q) \in E_{{\cal F}_q}$ for some  $K_q' \in {\cal F}_q$. In such a case, $Q_{({\cal K}, {\cal K}')} = K_q' \setminus K_q$. Therefore, by using Case (a) and the definition of the flow  $\Lambda^{\Phi^{q}}$ on $\Gamma_{{\cal F}_q}$,  one obtains
\begin{eqnarray}
\sum_{({\cal K}, {\cal K}') \in E_{\cal F} } \Lambda^{\Phi^{0}} ({\cal K}, {\cal K}') &=&  \sum_{({\cal K}, {\cal K}') \in E_{\cal F}^R } \Lambda^{\Phi^{0}} ({\cal K}, {\cal K}') \nonumber \\
                                                                                                                            &=&  \sum_{({\cal K}, {\cal K}') \in E_{\cal F}^R }  \Lambda^{\Phi^{M}} ({\cal R}_{\mu ({\cal K}) \setminus q}, {\cal R}_{\mu ({\cal K})} ) \sum_{i \in K_q' \setminus K_q} \lambda_i^{\Phi^{0}} ({\cal K}, { \cal K}')  \nonumber \\
                                                                                                                             &=&  \sum_{({\cal K}, {\cal K}') \in E_{\cal F}^R }  \Lambda^{\Phi^{M}} ({\cal R}_{\mu ({\cal K}) \setminus q}, {\cal R}_{\mu ({\cal K})} ) 
                                                                                                                             \Lambda^{\Phi^{q}} (K_q, K_q')     \nonumber \\
          &= &   \Lambda^{\Phi^{M}} ({\cal R}_{\mu ({\cal K}) \setminus q}, {\cal R}_{\mu ({\cal K})} ) \sum_{(K_q,  K_q') \in E_{{\cal F}_q }}  \Lambda^{\Phi^{q}} (K_q, K_q').              \nonumber                                                                                                 
   \end{eqnarray}                                                                                                                                                                                                           
Proceeding in a similar way for directed edges  $({\cal K}', {\cal K}) \in E_{\cal F}$, one gets
$$\sum_{({\cal K}', {\cal K}) \in E_{\cal F} } \Lambda^{\Phi^{0}} ({\cal K}', {\cal K})  =  \Lambda^{\Phi^{M}} ({\cal R}_{\mu ({\cal K}) \setminus q}, {\cal R}_{\mu ({\cal K})}) \sum_{ (K_q', K_q) \in E_{ {\cal F}_q }}  \Lambda^{\Phi^{q}} (K'_q, K_q).$$                                                                                                                                                                                                           
Thanks to the conservation constraints applied to the flow $\Lambda^{\Phi^{q}}$,
 $$  \sum_{(K_q', K_q) \in E^i_{\cal F} }  \Lambda^{\Phi^{q}} (K'_q, K_q) = \sum_{(K_q,  K_q') \in E^i_{{\cal F}_q } } \Lambda^{\Phi^{q}} (K_q, K_q') $$
one obtains
$$ \sum_{({\cal K}', {\cal K}) \in E_{\cal F} } \Lambda^{\Phi^{0}} ({\cal K}', {\cal K}) = \sum_{({\cal K}, {\cal K}') \in E_{\cal F} } \Lambda^{\Phi^{0}} ({\cal K}, {\cal K}'),$$             
 showing that the flow induced by $\Lambda^{\Phi^{0}}$ and passing through such a coalition profile ${\cal K}$ satisfies the conservation constraints.                                                                                                                                                                                                     
                                                                                                                                                                                                
From Cases (a), (b) and (c), conclude that  $\Lambda^{\Phi^{0}}: E_{\cal F} \longrightarrow \mathbb{R}$ defined as in (\ref{flow-2step}) is a flow. It remains to prove that this flow is unitary.

$$\sum_{(\emptyset_M, {\cal K}) \in E_{\cal F} } \Lambda^{\Phi^{0}} (\emptyset_M, {\cal K}) =  \sum_{q \in M }  \Lambda^{\Phi^{M}} (\emptyset_M, {\cal R}_{q}) \sum_{(\emptyset, K_q ) \in E_{{\cal F}_q}} \Lambda^{\Phi^{q}} (\emptyset, K_q).$$
Since  $\Lambda^{\Phi^{M}}$ and $\Lambda^{\Phi^{q}}$ are unitary flows,
$$\sum_{(\emptyset_M, {\cal K}) \in E_{\cal F} } \Lambda^{\Phi^{0}} (\emptyset, {\cal K}) = 1,$$
as desired.
\qed

According to Lemma \ref{Marg-two-step} and Theorem \ref{char-flowmethod}, a two-step flow method $\Phi^{0}$ on  ${\cal G}_{N,  {\cal P}, {\cal F}}$
defined through the two-step procedure (\ref{2-step-flow})-(\ref{2-step-flow1}) can be expressed as:
 \begin{eqnarray}
\forall i\in N, \quad \Phi^{0}_i (N, v, {\cal P}, {\cal F})  &= & \sum_ {q \in M: \atop i \in P_q } \Phi_i^{q} (P_q, v_q, (P_q), {\cal F}_q),  \nonumber 
 \end{eqnarray} 
where
$$ \Phi_i^{q} (P_q, v_q, (P_q), {\cal F}_q) = \!\!\!\!\!\! \!\!\! \! \! \sum_{({\cal K}, { \cal K'}) \in E^R_{\cal F} \cap E^i_{{\cal F}}: \atop q_{({\cal K}, { \cal K'})} = q} \Lambda^{\Phi^M}({\cal R}_{ \mu ({\cal K}') \setminus q},{\cal R}_{ \mu ({\cal K}')} )  \lambda^q_i(K_q, K'_q)   
\bigl (v({\cal K}') - v({\cal K})\bigr).$$
For each $q$ and each directed edge $(K_q, K_q') \in E_{{\cal F}_q}$, the sum of the coefficients is equal to the flow induced by $\Phi^q$ passing through the directed edges $(K_q, K_q')$, that is,   
$$ \sum_{i \in K_q' \setminus K_q}  \lambda_i^{\Phi^{q}} (K_q, K_q') = \Lambda^{\Phi^q} (K_q, K_q').$$

Among the family of flow methods are those based on maximal paths, in the spirit of Shapley's procedure, where the grand coalition $N$ is formed step by step from a linear order on the agents in the set $N$. Inspired by this idea, Owen \cite{Owen1977} and Albizuri et al. \cite{Albizuri2006a} (implicitly) construct their value from certain maximal paths and unitary flows on these paths. 

In our context, consider a maximal path $W^{} = (\emptyset_M, \ldots, (\{1\}, \ldots, \{m\}))$ in the directed hypercube $\Gamma_{{\cal F}^M}$ of dimension $m$.
There are exactly $m !$ maximal paths in $\Gamma_{{\cal F}^M}$. Denote by $P_{\Gamma_{\cal F}}$ the set of these $m !$ maximal paths.
Define the unitary flow $\Lambda^{M}_W: E_{{\cal F}^M} \longrightarrow \mathbb{R}$ such that  $\Lambda^{M}_W ({\cal R}, {\cal R}') = 1$
if the pair $({\cal R}, {\cal R}')$ lies on the path and $\Lambda^{M}_W ({\cal R}, {\cal R}') = 0$ otherwise.
Next, consider, 
$$\Lambda^{M} = \frac{1}{m !} \sum_{W \in P_{\Gamma_{\cal F}} } \Lambda^{M}_W.$$
Note that $\Lambda^{M}$ is still a unitary flow on  $\Gamma_{{\cal F}^M}$ since  the set of flows on a digraph forms a linear subspace of $\mathbb{R}^{|E|}$ where $|E|$ is the number of directed edges of the digraph. Therefore,
$$\forall S \subseteq M, \forall q \in S, \quad \Lambda^{M} (\mathcal{R}_{S \setminus q},  \mathcal{R}_{S} )= \frac{(m - s) ! (s - 1)!}{m !},$$   
where $(s - 1) !$ is the number of paths from $\emptyset_M$ to  $\mathcal{R}_{S \setminus q}$, and $(m - s) !$ is the number of paths from $\mathcal{R}_{S}$ to $(\{1\}, \ldots, \{m\})$. 
Consequently, the flow method $\Phi^M$ is the extension of the Shapley value from TU-games to the upper-games $(M, v_M, P_M, {\cal F}_M)$ given by:
\begin{equation} \label{shapley}
\forall q \in M, \quad \Phi_q^M (M, v_M, P_M, {\cal F}_M ) = \sum_{ S \subseteq M: \atop q \in S}  \frac{(m - s) ! (s - 1)!}{m !} \bigl ( v_M(\mathcal{R}_S) - v_M(\mathcal{R}_{S \setminus q}) \bigr).
\end{equation}

For each $q \in M$, proceed as above in each digraph $\Gamma_{{\cal F}_q}$. Denote by $\ell_p$ the number of maximal paths in  $\Gamma_{{\cal F}_q}$ and, for each $(K_q, K_q') \in E_{{\cal F}_q}$,  $\gamma_q (\emptyset, K_q)$ is the number of paths from $\emptyset$ to $K_q $ and  $\gamma_q (K'_q, P_q)$ is the number of paths from $K_q'$ to $P_q$. This results in  the unitary flow $\Lambda^{q}: E_{{\cal F}_q} \longrightarrow \mathbb{R}$ defined as:
$$(K_q, K_q') \in E_{{\cal F}_q}, \quad  \Lambda^{q} (K_q, K_q') = \frac{\gamma_q (\emptyset, K_q) \gamma_q (K_q', P_q)}{\ell_p}.$$
Assuming that the flow  $\Lambda^{q} (K_q, K_q')$ is distributed equally among the agents in $K_q' \setminus K_q$, the resulting
flow method $\Phi^q (P_q, (P_q), {\cal F}_q)$ is a Shapley-like value where all maximal paths have equal weight and where the contribution from the feasible coalition $K_q$ to its cover $K_q'$ is shared equally among the agents in $K'_q \setminus K_q$. In other words, $\Phi^q$ can be viewed as the extension of the Shapley value from TU-games to TU-games with set system (see, Aguilera et al., \cite{Aguilera2010}, p.131):
\begin{equation} \label{ext-Shapley-q}
\forall i \in P_q,   \,\,  \Phi_i^q (P_q, (P_q), {\cal F}_q) = \!\!\! \sum_{(K_q, K_q') \in \Gamma_{{\cal F}_q}: \atop i \in K_q' \setminus K_q }    \frac{\gamma_q (\emptyset, K_q) \gamma_q (K_q', P_q)}{\ell_p | K_q' \setminus K_q|} \bigl ( v_q (K'_q) - v_q(K_q)  \bigr).
\end{equation}
Consequently, the  two-step flow method $\Phi^0$ on ${\cal G}_{N,  {\cal P}, {\cal F}}$ is given by:
$$\forall i \in N, \quad \Phi^{0}_i (N, v, {\cal P}, {\cal F}) = $$
\begin{equation} \label{G-Albi}
\sum_ {q \in M: \atop i \in P_q } \sum_{S \subseteq M: \atop q \in S}\!\!\! \sum_{(K_q, K_q') \in E_{{\cal F}_q}}\frac{(m - s) ! (s - 1)!}{m !}  \frac{\gamma_q (\emptyset, K_q) \gamma_q (K_q', P_q)}{\ell_p | K_q' \setminus K_q|}
\bigl (v({\cal K}_{S, K_q'}) - v({\cal K}_{S, K_q} )\bigr).
 \end{equation} 
From (\ref{shapley}) and  (\ref{ext-Shapley-q}), $\Phi^{0}$ is computed from the two-step procedure (\ref{2-step-flow})-(\ref{2-step-flow1})  where at each step the Shapley-like value is applied. Equivalently, the two-step flow $\Lambda^{\Phi^0}$ associated with $\Phi^{0} $ is given, for each relevant edge, by
 \begin{eqnarray} \label{flow-Shapley-1}
 \Lambda^{\Phi^0} ({\cal K}_{S, K_q}, {\cal K}_{S, K'_q}) &=& \Lambda^{M} ( {\cal R}_{S \setminus q}, {\cal R}_S)  \Lambda^{q} (K_q, K_q') \nonumber \\
 & = & \frac{(m - s) ! (s - 1)!}{m !}  \frac{\gamma_q (\emptyset, K_q) \gamma_q (K_q', P_q)}{\ell_p}.
\end{eqnarray} 

To illustrate the graphical construction of the two-step flow method defined in (\ref{G-Albi}), consider the following example.
\begin{example} Consider the subdigraph represented in Figure \ref{relevant-graph2}. It is the subdigraph of $\Gamma_{{\cal F}}$ (see Figure \ref{fig2-1})  induced by the set of relevant directed edges $E^R_{{\cal F}}$. Figure \ref{Hyper-product} adds to this subdigraph of $\Gamma_{{\cal F}}$ the hypercube of dimension $m = 2$ corresponding to $\Gamma_{{\cal F}^M}$, where the relevant directed edges are in dotted lines.  
Therefore, for each (dotted) directed edge $(\mathcal{R}_{S \setminus q},  \mathcal{R}_{S} )$,  the flow is given by:
$$\Lambda^{M} (\mathcal{R}_{S \setminus q},  \mathcal{R}_{S} )= \frac{(m - s) ! (s - 1)!}{m !}  = {\color {blue} \frac{1}{2}}.$$
Next, consider the directed edge $((\{2, 3\}, \{3, 4, 5\}), (\{1, 2, 3\}, \{3, 4, 5\}))$  of $\Gamma_{\cal F}$. In $\Gamma_{{\cal F}_1}$ the number $\ell_1$ of maximal paths is equal to 2,
the number $\gamma_1(\emptyset, \{2, 3\})$ of directed paths  from $\emptyset$ to $\{2, 3\}$ is equal to 1, and the number $\gamma_1(\{1, 2, 3\}, \{1, 2, 3\})$ of directed paths  from $\{1, 2, 3\}$ to $\{1, 2, 3\}$ is equal to 1, so that
$$  \frac{\gamma_1 (\emptyset,  \{2, 3\}) \gamma_1 (\{1, 2, 3\}, \{1, 2, 3\})}{\ell_1} =   \frac{1 \times 1}{2} =  \frac{1}{2}.$$
It follows that the flow on the directed edge  $((\{2, 3\}, \{3, 4, 5\}), (\{1, 2, 3\}, \{3, 4, 5\}))$ is given by
$$\Lambda^{M} ((\emptyset, \{2\}), (\{1\}, \{2\})) \frac{\gamma_1 (\emptyset,  \{2, 3\}) \gamma_1 (\{1, 2, 3\}, \{1, 2, 3\})}{\ell_1} =  {\color {blue}\frac{1}{2}} \times \frac{1}{2} =   {\color {red}\frac{1}{4}}.$$
As Agent 1 is the only agent to join $\{2, 3\} $ in $P_1$, the above flow is allocated to this agent.
\begin{figure} [h] 
\begin{center}
\begin{tikzpicture} [every node/.style={scale = 0.6}, scale = 0.6]

  \node (A) at (0,0) {$(\emptyset, \emptyset)$};
  \node (B) at (5,0) {$(\{2, 3\}, \emptyset)$};
  \node (C) at (2.5, 1.5) {$(\{1\}, \emptyset)$};
  \node (D) at (7.5, 1.5) {$(\{1, 2, 3\}, \emptyset)$};

  \draw[->] (A) -- (B);
  \draw[->] (B) -- (D);
  \draw [->] (C) -- (D);
  \draw [->] (A) -- (C);

 \node (E) at (0, 3) {$(\emptyset, \{3, 4\})$};
\node (H) at (7.5, 4.5) {$(\small \{1, 2, 3\}, \{3, 4\})$};  
  
  
  \node (I) at (0, 6) {$(\emptyset, \{3, 4, 5\})$};
\node (J) at (5, 6) {(\{2, 3\}, \{3, 4, 5\})};
\node (K) at (2.5, 7.5) {(\{1\}, \{3, 4, 5\})};
\node (L) at (7.5, 7.5) {(\{1, 2, 3\}, \{3, 4, 5\})};

\draw[->, dashed] (-0.5 ,0)  arc  (225:135:4);
\draw [->, dashed] (I) -- (L);
\draw [->, dashed] (A) -- (D);
\draw[->, dashed] (9 ,1.5)  arc  (-45:45:4);

 \node (M) at (3, 6.8 ) {\small \color{blue} 0.5  };
 \node (N) at (6.5, 6.5) {\small \color{red} 0.25 };

   \draw [->] (I) -- (J);
 \draw [->] (I) -- (K);
 \draw [->] (J) -- (L);
 \draw [->] (K) -- (L);  

\draw [->] (A) -- (E);
\draw [->] (E) -- (I);

  
  
   \draw [->] (D) -- (H);
  \draw [->] (H) -- (L);  

\end{tikzpicture}  
\end{center} 
\caption{The subdigraph of $\Gamma_{{\cal F}}$ induced by $E_{\cal F}^R$ and the hypercube $\Gamma_{{\cal F}^M}$  of dimension 2.}
\label{Hyper-product}
\end{figure}
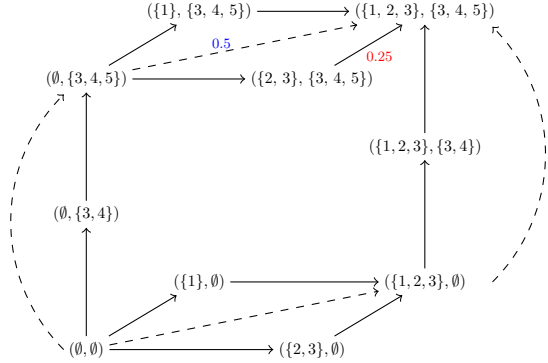
Proceeding as above, Figure \ref{flow-relevant-3} gives the flow (the values in red) on $\Gamma_{{\cal F}^3}$, that is the flow passing through the relevant directed edges involving agent 3. 
Therefore, according to (\ref{G-Albi}), the flow assigned to agent 3 is given by the values in brackets.
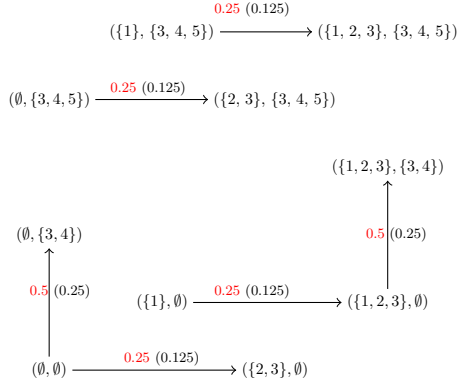
\begin{figure}
\begin{center}
\begin{tikzpicture}  [every node/.style={scale = 0.6}, scale = 0.6]  
  \node (A) at (9,0) {$(\emptyset, \emptyset)$};
  \node (B) at (14,0) {$(\{2, 3\}, \emptyset)$};
  \node (C) at (11.5, 1.5) {$(\{1\}, \emptyset)$};
  \node (D) at (16.5, 1.5) {$(\{1, 2, 3\}, \emptyset)$};

  \draw[->] (A) -- (B);
  \draw [->] (C) -- (D);

 \node (E) at (9, 3) {$(\emptyset, \{3, 4\})$};
\node (H) at (16.5, 4.5) {$(\small \{1, 2, 3\}, \{3, 4\})$};  
  

\node (O) at (11.5, 0.25) {\small {\color{red} 0.25}  (0.125)};
\node (P) at (13.5, 1.75) {\small {\color{red} 0.25} (0.125)};
\node (Q) at (11.2, 6.25)  {\small {\color{red} 0.25} (0.125)};
\node (R) at (13.5, 8)  {\small {\color{red} 0.25} (0.125)};

\node (S) at (9.25, 1.75) {\small {\color{red} 0.5} (0.25)};
\node (T) at (16.70, 3) {\small    {\color{red} 0.5}  (0.25)};
  
  \node (I) at (9, 6) {$(\emptyset, \{3, 4, 5\})$};
\node (J) at (14, 6) {(\{2, 3\}, \{3, 4, 5\})};
\node (K) at (11.5, 7.5) {(\{1\}, \{3, 4, 5\})};
\node (L) at (16.5, 7.5) {(\{1, 2, 3\}, \{3, 4, 5\})};

   \draw [->] (I) -- (J);
 \draw [->] (K) -- (L);  

\draw [->] (A) -- (E);

  
  
   \draw [->] (D) -- (H);
 
\end{tikzpicture}
\end{center}
\caption{In red, the flow passing through the relevant directed edges of agent 3; in brackets, the flow assigned to agent 3.  For vertical directed edges, the flow is uniformly shared between agents 3 and 4 and for the other edges the flow  is shared between agents 2 and 3.  }
\label{flow-relevant-3}
\end{figure}
\end{example}

\begin{remark}
 The flow $\Lambda^{\Phi^0}$ in (\ref{flow-Shapley-1}) can also be directly constructed  using 
 the subdigraph of $\Gamma_{{\cal F}}$  induced by the set of relevant directed edges $E^R_{{\cal F}}$. 
 This subdigraph has  $m ! \,  \ell_1 \, \ell_2 \ldots \, \ell_m$ maximal paths. Let $ P_{\Gamma_{\cal F}}^R$  be the set of such relevant maximal paths. 
 For any  path $W  \in  P_{\Gamma_{\cal F}}^R$,  define the unitary flow $\Lambda^R_{W}$ as above. 
Then one obtains
$$\Lambda^{\Phi^0} =  \frac{1}{m !\,  \ell_1 \, \ell_2 \ldots \, \ell_m}  \sum_{W  \in   P_{\Gamma_{\cal F}}^R} \Lambda^R_{W}.$$
\end{remark}

\begin{example}  \label{ExCGAS}
Another specific instance of the two-step flow method given in (\ref{G-Albi}) is the situation where, for each $q \in M$, $(P_q, {\cal F}_q)$ is either a convex geometry or an augmenting system containing $P_q$. In this case, the two-step flow method given in (\ref{G-Albi}) is as follows.
If $(P_q, {\cal F}_q)$ is a convex geometry,
$\Phi^q (P_q, v_q, (P_q), {\cal F}_q)$ is the Shapley value for convex geometry  as introduced in Bilbao \cite{Bilbao1998} and Bilbao and Edelman \cite{Bilbao2000}; if 
$(P_q, {\cal F}_q)$ is an augmenting system, $\Phi^q (P_q, v_q, (P_q), {\cal F}_q)$ is the Shapley value for augmenting systems  introduced in Bilbao and Ord\'o\~{n}ez \cite{Bilbao_Ordonez}; and $\Phi^M$ is the extension of the Shapley value from TU-games to the upper-games given by (\ref{shapley}).
In a convex geometry and in an augmenting system, for each pair $(K_q, K'_q) \in E_{{\cal F}_q}$, $q \in M$, $K'_q \setminus K_q = \{i\}$ for $i \in N$. Consequently, the flow
$\Lambda^q (K_q, K'_q)$ is fully assigned to this agent $i$.
\end{example}

\begin{remark}
Our framework is sufficient general to include other specific cases such as situations where the set systems are derived from a precedence constraints relation (Faigle and Kern \cite{F1992}). In the context of cooperative games with precedent constraints, Faigle and Kern \cite{F1992} define a Shapley-like value that can be used in the same way as in Example \ref{ExCGAS}. See also
 Algaba et al. \cite{ABD2017} for values in the context of cooperative games with precedent constraints.
\end{remark}

%
%

\section{Characterization of two-step flow methods}  \label{Section-Two-step-FM-Chara}

We say that a flow $\Lambda^{0}$ on the product digraph $\Gamma_{\cal F}$   is a {\bf two-step flow} if the flow is constructed from flows $\Lambda^q$ on the digraphs $\Gamma_{{\cal F}_q}$, $q \in M$, and from a flow $\Lambda^{M}$ on the directed hypercube $\Gamma_{{\cal F}^M}$  in the following way:
 \begin{eqnarray}  \label{2-step-flow2}
1. & & \forall ({\cal K}, {\cal K}') = ({\cal K}_{\mu ({\cal K}'), K_q}, {\cal K}'_{\mu ({\cal K}'), K'_q}) \in E_{\cal F}^{R}, \, q \in \mu ({\cal K}'),  \nonumber \\
   & & \Lambda_{} ({\cal K}, {\cal K}') = \Lambda^{M} ({\cal R}_{\mu ({\cal K}') \setminus q}, {\cal R}_{ \mu ({\cal K}')}) \Lambda^{q} (K_q, K'_q); \nonumber \\
2.& & \Lambda_{}  ({\cal K}, {\cal K}') = 0, \mbox{ otherwise}.
\end{eqnarray}
By Theorem \ref{char-flowmethod}, a two-step flow method $\Phi^{0}$ on ${\cal G}_{N, {\cal P}, {\cal F}}$ induces a two-step flow $\Lambda^{\Phi^{0}}$ on the product digraph $\Gamma_{\cal F}$. Reciprocally,  from any two-step flow $\Lambda^{0}$ on the product digraph $\Gamma_{\cal F}$, one can construct a two-step flow method $\Phi^{0}$ on ${\cal G}_{N, {\cal P}, {\cal F}}$. It turns out that the two-step flow methods can be characterized by two axioms for flows on $\Gamma_{\cal F}$. The  first axiom stipulates that the flow passing through non-relevant directed edges of the product digraph is null. The second axiom expresses a principle of flow proportionality between relevant directed edges of the product digraph $\Gamma_{\cal F}$.  \\

\noindent {\bf Null flow for non-relevant directed edges}. A flow $\Lambda$ on $\Gamma_{\cal F}$ satisfies null flow for non-relevant edges if for each non-relevant directed edge $({\cal K},{\cal K}') \in E_{\cal F}\setminus E_{\cal F}^R$, it holds that $\Lambda_{} ( {\cal K},{\cal K}') = 0$. \\

\noindent {\bf Flow proportionality}. A flow $\Lambda$ on $\Gamma_{\cal F}$ satisfies flow proportionality if for each $q \in M$, each pair of relevant directed edges $\{(K_q, K_q'), (L_q, L_q')\} \subseteq E_{{\cal F}_q}$ of $\Gamma_{{\cal F}_q}$ and each pair $\{S, S'\} \subseteq M$ such that $q \in S \cap S'$, the following proportionality principle holds:
 $$\Lambda_{} ({\cal K}_{S, K_q}, {\cal K}_{S, K'_q} ) \Lambda_{} ({\cal K}_{S', L_q}, {\cal K}_{S', L'_q} )= \Lambda_{} ({\cal K}_{S', K_q}, {\cal K}_{S', K'_q} ) \Lambda_{} ({\cal K}_{S, L_q}, {\cal K}_{S, L'_q} ).$$

\begin{theorem}\label{char-twostepflow1}
A  flow $\Lambda$ on the product digraph $\Gamma_{\cal F}$ is a two-step flow as in (\ref{2-step-flow2})  if and only if it satisfies Null flow for non-relevant directed edges and Flow proportionality.
\end{theorem} 

The proof of Theorem \ref{char-twostepflow1} relies on an intermediary result. 

Pick any  $S \subseteq M$, any $q \in S$, and the relevant coalition profile ${\cal K}_S$. Define by $({\cal F}^{S\setminus q, S}, E_{
{\cal F}^{S\setminus q, S}})$ the subdigraph of $\Gamma_{\cal F}$ induced 
by the subset ${\cal F}^{S\setminus q, S} \subseteq {\cal F}$ of coalition profiles between ${\cal K}_{S \setminus q}$  and  ${\cal K}_S$,
$${\cal F}^{S\setminus q, S} = \bigl\{{\cal K} \in {\cal F}:  {\cal K}_{S \setminus q}  \subseteq^M {\cal K} \subseteq^M {\cal K}_S \bigr\}.$$
Note that each element of ${\cal F}^{S\setminus q, S}$ is a relevant coalition profile.

\begin{lemma} \label{Null-flow-M}
Let $\Lambda$ be a unitary flow on $\Gamma_{\cal F}$ satisfying Null flow for non-relevant directed edges.
Then, the following two points hold.

\begin{enumerate}
\item The unitary flow $\Lambda$ on $\Gamma_{\cal F}$ induces a flow on the subdigraph $({\cal F}^{S\setminus q, S}, E_{
{\cal F}^{S\setminus q, S}})$;
\item Define the function  $\Lambda^M$ applied to the directed hypercube $\Gamma_{{\cal F}^M}$ as
\begin{equation} \label{flowMprod}
\forall S \subseteq M \setminus \emptyset, \forall q \in S, \quad 
\Lambda^M ({\cal R}_{S \setminus q}, {\cal R}_S) = \sum_{ K_q \in {\cal F}_q: \atop (\emptyset, K_q) \in E_{{\cal F}_q} }\Lambda_{} ({\cal K}_{S\setminus q}, {\cal K}_{S, K_q}), 
\end{equation}
or equivalently, 
$$\Lambda^M ({\cal R}_{S \setminus q}, {\cal R}_S)= \sum_{{\cal K} \subseteq^M {\cal K}_{S}: \atop ({\cal K}_{S \setminus q}, {\cal K}) \in E_{\cal F}}\Lambda_{} ({\cal K}_{S\setminus q}, {\cal K}). $$
Then, $\Lambda^M$ is a unitary flow on $\Gamma_{{\cal F}^M}$.

\end{enumerate} 
\end{lemma}
{\it Proof} \textbf{Point 1}.  Pick any ${\cal K} \in {\cal F}^{S\setminus q, S}$ different from ${\cal K}_{S \setminus q}$  and ${\cal K}_{S} $ and any directed edge $({\cal K}, {\cal K}') \in E_{\cal F}$. Then $({\cal K}, {\cal K}') \in E_{
{\cal F}^{S\setminus q, S}}$ if and only if  $ {\cal K}'$ is relevant.  By Null flow for non-relevant directed edges, it follows that
$$\sum_{{\cal K}' \in {\cal F}: \atop ({\cal K}, {\cal K}') \in E_{\cal F}}\Lambda_{ } ({\cal K}, {\cal K}') = \sum_{{\cal K}' \in {\cal F}: \atop ({\cal K}, {\cal K}') \in E_{
{\cal F}^{S\setminus q, S}}}\Lambda_{ } ({\cal K}, {\cal K}').$$  
In the same way,
$$\sum_{{\cal K}' \in {\cal F}: \atop ({\cal K}', {\cal K}) \in E_{\cal F}}\Lambda_{} ({\cal K}', {\cal K}) = \sum_{{\cal K}' \in {\cal F}: \atop ({\cal K}', {\cal K}) \in E_{{\cal F}^{S\setminus q, S}}}\Lambda_{} ({\cal K}', {\cal K}).$$
Since  $\Lambda_{ }$ is a flow on $\Gamma_{\cal F}$, it satisfies the conservation constraints so that
$$\sum_{{\cal K}' \in {\cal F}^{S\setminus q, S}: \atop ({\cal K}, {\cal K}') \in E_{
{\cal F}^{S\setminus q, S}}}\Lambda_{  } ({\cal K}, {\cal K}') =  \sum_{{\cal K}' \in {\cal F}^{S\setminus q, S} \atop ({\cal K}', {\cal K}) \in E_{{\cal F}^{S\setminus q, S}}}\Lambda_{  } ({\cal K}', {\cal K}),$$
which means that the conservation constraints are also satisfied for each vertex of $({\cal F}^{S\setminus q, S}, E_{
{\cal F}^{S\setminus q, S}})$ different from ${\cal K}_{S \setminus q}$ and ${\cal K}_{S}$. Thus, by definition of a flow,   
$\Lambda_{ }$ induces a flow on the subdigraph $({\cal F}^{S\setminus q, S}, E_{
{\cal F}^{S\setminus q, S}})$. 

{\bf Point 2}.  Consider the cut  of $({\cal F}^{S\setminus q, S}, E_{
{\cal F}^{S\setminus q, S}})$ formed by the coalition profile ${\cal K}_{S \setminus q}$ and the coalition profiles ${\cal K}$ such that $({\cal K}_{S \setminus q}, {\cal K}) \in  E_{
{\cal F}^{S\setminus q, S}}$. In a similar way, consider the cut  of $({\cal F}^{S\setminus q, S}, E_{
{\cal F}^{S\setminus q, S}})$ formed by the coalition profile ${\cal K}_{S}$ and the coalition profiles ${\cal K}$ such that $({\cal K}, {\cal K}_{S}) \in  E_{
{\cal F}^{S\setminus q, S}}$. As $\Lambda_{ }$ induces a flow on the subdigraph $({\cal F}^{S\setminus q, S}, E_{
{\cal F}^{S\setminus q, S}})$ by Point 1, and that
the value of a flow is equal to the value of any cut, one obtains:
$$\sum_{{\cal K} \in {\cal F}^{S\setminus q, S}: \atop ({\cal K}_{S \setminus q}, {\cal K}) \in E_{
{\cal F}^{S\setminus q, S}}}\Lambda_{ } ({\cal K}_{S \setminus q}, {\cal K}) =  \sum_{{\cal K} \in {\cal F}^{S\setminus q, S}: \atop ({\cal K}, {\cal K}_{S}) \in E_{{\cal F}^{S\setminus q, S}   }}\Lambda_{ } ({\cal K}, {\cal K}_S),$$
or equivalently
$$\sum_{{\cal K} \subseteq^M {\cal K}_{S }: \atop ({\cal K}_{S \setminus q}, {\cal K}) \in E_{
{\cal F}}}\Lambda_{ } ({\cal K}_{S \setminus q}, {\cal K}) =  \sum_{{\cal K} \supseteq^M {\cal K}_{S \setminus q}:  \atop ({\cal K}, {\cal K}_{S}) \in E_{{\cal F}}}\Lambda_{ } ({\cal K}, {\cal K}_S).$$
By definition (\ref{flowMprod}) of $\Lambda^M$, the above quantity is equal to 
$\Lambda^M ({\cal R}_{S \setminus q}, {\cal R}_{S}) $.
Next, consider the sum of the quantities $\Lambda^M ({\cal R}_{S \setminus q}, {\cal R}_{S}) $  over all $q \in S$,
\begin{eqnarray}
\sum_{q \in S} \Lambda^M ({\cal R}_{S \setminus q}, {\cal R}_S)  &=& \sum_{q \in S} \sum_{{\cal K} \subseteq^M {\cal K}_{S }: \atop ({\cal K}_{S \setminus q}, {\cal K}) \in E_{
{\cal F}}}\Lambda_{ } ({\cal K}_{S \setminus q}, {\cal K})  \nonumber \\
&=& \sum_{q \in S}  \sum_{{\cal K} \supseteq^M {\cal K}_{S \setminus q}:  \atop ({\cal K}, {\cal K}_{S}) \in E_{{\cal F}}}\Lambda_{ } ({\cal K}, {\cal K}_S) \nonumber \\
&=&  \sum_{({\cal K}, {\cal K}_{S}) \in E_{\cal F}} \Lambda_{ } ({\cal K}, {\cal K}_{S}). \nonumber
\end{eqnarray} 
In a similar way,
$$\sum_{q \in M \setminus S} \Lambda^M ({\cal R}_{S}, {\cal R}_{S \cup q}) =   \sum_{({\cal K}_S, {\cal K}) \in E_{\cal F}} \Lambda_{ } ({\cal K}_S, {\cal K}_{}).$$
Since $\Lambda$ is a flow on $\Gamma_{\cal F}$, the conservation constraint leads to
$$\sum_{q \in S} \Lambda^M ({\cal R}_{S \setminus q}, {\cal R}_S) = \sum_{q \in M \setminus S} \Lambda^M ({\cal R}_{S}, {\cal R}_{S \cup q}),$$
which ensures that $\Lambda^M$ is a flow on $\Gamma_{{\cal F}^M}$. It remains to verify that this flow is unitary. 
$$\sum_{q \in M} \Lambda^M (\emptyset_{M}, \{q\})  =  \sum_{q \in M} \sum_{{\cal K} \subseteq^M {\cal K}_{\{q\}}, \atop (\emptyset_M,{\cal K}) \in E_{{\cal F}}  } \Lambda (\emptyset_{M}, {\cal K}) = \sum_{(\emptyset_M,{\cal K}) \in E_{{\cal F}}  } \Lambda (\emptyset_{M}, {\cal K}) = 1,$$
where the last equality comes from the fact that the flow $\Lambda$ is unitary. Therefore, the flow $\Lambda^M$  on $\Gamma_{{\cal F}^M}$ is unitary.
\qed

{\it Proof} ({\bf of Theorem \ref{char-twostepflow1}}):
Assume first that $\Lambda$ is a two-step flow on $\Gamma_{\cal F}$ constructed from flows $\Lambda^q$ applied to the digraphs $\Gamma_{{\cal F}_q}$, $q \in M$, and from a flow $\Lambda^{M}$ applied to the directed hypercube $\Gamma_{{\cal F}^M}$ as in Points 1 and 2 above. From its definition, $\Lambda$ satisfies Null flow for non-relevant directed edges. Next, pick any $q\in M$,  any pair of directed edges $\{(K_q, K_q'), (L_q, L_q')\} \subseteq E_{{\cal F}_q}$ of $\Gamma_{{\cal F}_q}$ and any pair $\{S, S'\} \subseteq M$ such that $q \in S \cap S'$. By Point 1 of the definition of a two-step flow $\Lambda$ given in (\ref{2-step-flow2})
$$\Lambda_{} ({\cal K}_{S, K_q}, {\cal K}_{S, K_q'}) \Lambda_{} ({\cal K}_{S', L_q}, {\cal K}_{S', L'_q}), $$
 is equal to
$$ \Lambda^{M} ({\cal R}_{S\setminus q}, {\cal R}_{S}) \Lambda^{q} (K_q, K'_q) \Lambda^{M} ({\cal R}_{S'\setminus q}, {\cal R}_{S' }) \Lambda^{q} (L_q, L'_q),$$
expression that coincides with the definition of
$$\Lambda_{} ({\cal K}_{S', K_q}, {\cal K}_{S', K_q'}) \Lambda_{} ({\cal K}_{S, L_q}, {\cal K}_{S, L'_q}).$$
Thus, $\Lambda$ satisfies Flow proportionality for relevant edges.

Reciprocally,  assume that $\Lambda$ is a flow on $\Gamma_{{\cal F}}$ that satisfies Null flow for non-relevant directed edges and Flow proportionality. Let $\Lambda^M$ be the flow on the directed hypercube $\Gamma_{{\cal F}^M}$ as defined in the statement of Lemma \ref{Null-flow-M}. For each $q \in M$, the subset of directed edges
$$\bigl\{({\cal R}_{S \setminus q }, {\cal R}_{S}) \subseteq E_{{\cal F}^M}: S \subseteq  M, S \ni q \bigr\}, $$
 forms a cut of $\Gamma_{{\cal F}^M}$. By Lemma \ref{Null-flow-M},  $\Lambda^M$ is a unitary flow so that
$$\sum_{S \subseteq M: \atop S \ni q} \Lambda^M ({\cal R}_{S\setminus q}, {\cal R}_{S} )  = 1.$$
It follows that there is $S^* \ni q$ such that  $\Lambda^M ({\cal R}_{S^*\setminus q}, {\cal R}_{S^* } ) \not = 0$. Consider such a directed edge, and define $\Lambda^q$ on $E_{{\cal F}_q}$ as follows:
 \begin{equation} \label{def-lambda-q}
 \forall (K_q, K_q') \in E_{{\cal F}^q}, \quad \Lambda^q(K_q, K_q') = \frac{ \Lambda({\cal K}_{S^*, K_q}, {\cal K}_{S^*, K'_q}) }{\Lambda^M ({\cal R}_{S^* \setminus q}, {\cal R}_{S^*} )}.
 \end{equation}
We first show that $\Lambda^q$ is a unitary flow on the directed edges of  ${\cal F}_q$. Let $K_q \in {\cal F}_q \setminus \{\emptyset, P_q\}$. We have
\begin{eqnarray}
\sum_{K'_q \in  {\cal F}_q, \atop (K_q, K'_q) \in E_{{\cal F}_q}} \Lambda^q(K_q, K_q') &=&
 \sum_{K'_q \in  {\cal F}_q, \atop (K_q, K'_q) \in E_{{\cal F}_q}}
 \frac{ \Lambda({\cal K}_{S^*, K_q}, {\cal K}_{S^*, K'_q}) }{\Lambda^M ({\cal R}_{S^* \setminus q}, {\cal R}_{S^*} )} \nonumber \\
&= & \sum_{{\cal K}\in  {\cal F}, \atop ({\cal K}_{S^*, K_q}, {\cal K}) \in E_{{\cal F}}} \frac{ \Lambda({\cal K}_{S^*, K_q}, {\cal K}) }{\Lambda^M ({\cal R}_{S^*\setminus q}, {\cal R}_{S^*} )}, \nonumber 
\end{eqnarray}  
where the last equality comes from the fact that $\Lambda$ satisfies Null flow for non-relevant directed edges and
$({\cal K}_{S^*, K_q}, {\cal K})$ is relevant if and only if ${\cal K}$ can written as ${\cal K}_{S^*, K'_q}$ for  $(K_q, K'_q) \in E_{{\cal F}_q}$. In a similar way,
\begin{eqnarray}
\sum_{K'_q \in  {\cal F}_q, \atop (K'_q, K_q) \in E_{{\cal F}_q}} \Lambda^q(K'_q, K_q) &=& \sum_{K'_q \in  {\cal F}_q, \atop (K'_q, K_q) \in E_{{\cal F}_q}} \frac{ \Lambda({\cal K}_{S^* , K'_q}, {\cal K}_{S^*, K_q}) }{\Lambda^M ({\cal R}_{S^*\setminus q}, {\cal R}_{S^*} )} \nonumber \\
&= & \sum_{{\cal K}\in  {\cal F}, \atop ({\cal K}, {\cal K}_{S^*, K_q}) \in E_{{\cal F}}}
\frac{ \Lambda( {\cal K}, {\cal K}_{S^*, K_q}) }{\Lambda^M ({\cal R}_{S^*\setminus q}, {\cal R}_{S^*} )}. \nonumber
\end{eqnarray}  
Using the conservation constraints for $\Lambda$, one obtains
$$ \sum_{K'_q \in  {\cal F}_q, \atop (K_q, K'_q) \in E_{{\cal F}_q}} \Lambda^q(K_q, K_q')  =  \sum_{K'_q \in  {\cal F}_q, \atop (K'_q, K_q) \in E_{{\cal F}_q}} \Lambda^q(K'_q, K_q),$$ 
which shows that $\Lambda^q$ is a flow on the directed edges of $ \Gamma_{{\cal F}_q}$. On the other hand,
\begin{eqnarray}
\sum_{K_q \in {\cal F}_q, \atop (\emptyset, K_q) \in E_{{\cal F}_q} } \Lambda^q (\emptyset, K_q) &=& \sum_{K_q \in  {\cal F}_q, \atop (\emptyset, K_q) \in E_{{\cal F}_q}} \frac{ \Lambda({\cal K}_{S^* \setminus q}, {\cal K}_{S^*, K_q}) }{\Lambda^M ({\cal R}_{S^* \setminus q}, {\cal R}_{S^*} )} \nonumber \\
&= & \frac{\Lambda^M ({\cal R}_{S^*\setminus q}, {\cal R}_{S^*} )}{\Lambda^M ({\cal R}_{S^* \setminus q}, {\cal R}_{S^*} )} \nonumber \\
&= & 1, \nonumber 
\end{eqnarray}
where the second equality comes from the definition of $\Lambda^M$ as in the statement of Lemma \ref{Null-flow-M}. 
This ensures that $\Lambda^q$ is unitary. It remains to show that $\Lambda$ can be expressed as a two-step flow using
the flow $\Lambda^M$ on $\Gamma_{{\cal F}^M}$ and the flows $\Lambda^q$ on $\Gamma_{{\cal F}_q}$, $q \in M$.
Consider any $q \in M$, any directed edge  $(K_q, K'_q) \in E_{{\cal F}_q}$, any $S \subseteq M$ such that $S \ni q$, and the coalition $S^*\ni q$ used in definition (\ref {def-lambda-q}) of $\Lambda^q$.
For $L_q \in {\cal F}_q$ such that $(\emptyset, L_q) \in  E_{{\cal F}_q}$, Flow proportionality indicates that

$$\Lambda_{} ({\cal K}_{S, K_q}, {\cal K}_{S, K'_q} ) \Lambda_{} ({\cal K}_{S^* \setminus q}, {\cal K}_{S^*, L_q} )\!\! =\!\! \Lambda_{} ({\cal K}_{S^* , K_q}, {\cal K}_{S^*, K'_q} ) \Lambda_{} ({\cal K}_{S \setminus q}, {\cal K}_{S, L'_q} ).$$ 

Using definition (\ref {def-lambda-q}) of $\Lambda^q$, the above equality rewrites
\begin{eqnarray}
& &\Lambda_{} ({\cal K}_{S, K_q}, {\cal K}_{S, K'_q} ) \Lambda^q(\emptyset, L_q)\Lambda^M ({\cal R}_{S^* \setminus q}, {\cal R}_{S^*} )  \nonumber \\ 
&&   \nonumber \\
& =& \Lambda^q(K_q, K_q')\Lambda^M ({\cal R}_{S^* \setminus q}, {\cal R}_{S^*} )\Lambda_{} ({\cal K}_{S \setminus q}, {\cal K}_{S, L_q} ), \nonumber
 \end{eqnarray}
that is,
$$\Lambda_{} ({\cal K}_{S, K_q}, {\cal K}_{S, K'_q} ) \Lambda^q(\emptyset, L_q) =     \Lambda^q(K_q, K_q')\Lambda_{} ({\cal K}_{S\setminus q}, {\cal K}_{S, L_q} ).$$
Summing over all directed edges $(\emptyset, L_q) \in E_{{\cal F}_q}$, one obtains
$$\Lambda_{} ({\cal K}_{S, K_q}, {\cal K}_{S, K'_q} ) \!\!\!\! \sum_{L_p \in {\cal F}_q, \atop (\emptyset, L_q) \in E_{{\cal F}_q}} \Lambda^q(\emptyset, L_q) =  \Lambda^q(K_q, K_q') \!\! \!\!\sum_{L_p \in {\cal F}_q, \atop (\emptyset, L_q) \in E_{{\cal F}_q}} \Lambda_{} ({\cal K}_{S \setminus q}, {\cal K}_{S, L_q} ).$$
As $\Lambda^q $ is unitary,
 $$\sum_{L_p \in {\cal F}_q, \atop (\emptyset, L_q) \in E_{{\cal F}_q}} \Lambda^q(\emptyset, L_q) = 1,$$
 and by definition of $\Lambda^M$,
 $$\Lambda^M ({\cal R}_{S \setminus q}, {\cal R}_{S} ) = \sum_{L_p \in {\cal F}_q, \atop (\emptyset, L_q) \in E_{{\cal F}_q}} \Lambda_{} ({\cal K}_{S \setminus q}, {\cal K}_{S, L_q} ),$$
 which leads to the desired result
 $$\Lambda_{} ({\cal K}_{S, K_q}, {\cal K}_{S, K'_q} ) = \Lambda^M ({\cal R}_{S\setminus q}, {\cal R}_{S} ) \Lambda^q(K_q, K_q').$$ 
\qed

As underlined above, from any two-step flow $\Lambda^{0}$ on the product digraph $\Gamma_{\cal F}$, one can construct a two-step flow method $\Phi^{0}$ on ${\cal G}_{N, {\cal P}, {\cal F}}$. Therefore, the combination of Theorem \ref{char-flowmethod} with  Theorem \ref{char-twostepflow1} leads to a characterization of the two-step flow methods among the flow methods in ${\cal G}_{N, {\cal P}, {\cal F}}$ in terms of the axioms of Null flow for non-relevant directed edges and  Flow proportionality on the product digraph $\Gamma_{\cal F}$.

\begin{theorem}\label{char-twostepflow}
 Let $\Phi^{0}$ be a flow method on ${\cal G}_{N, {\cal P}, {\cal F}}$ and let $\Lambda^{\Phi^{0}}$ be the induced flow on $\Gamma_{\cal F}$. Then, $\Phi^{0}$ is a two-step flow method as defined by the procedure (\ref{2-step-flow})-(\ref{2-step-flow1}) if and only if the flow $\Lambda^{\Phi^{0}}$  satisfies Null flow for non-relevant directed edges  and Flow proportionality for relevant edges on the product digraph $ \Gamma_{\cal F}$. 
\end{theorem} 

In the remaining part of this section,
we focus on the situation where cooperation among the agents is not restricted. In other words, we examine the specific case where all coalitions are feasible, that is ${\cal F}_q = 2^{P_q}$ for each $q \in M$. Denote by $2^{\cal P}$ the set of all coalition profiles and by
${\cal G}_{N, {\cal P},  2^{\cal P}} \subseteq {\cal G}_{N, {\cal P}, {\cal F}}$ the subdomain formed by the games $ (N, v, {\cal P}, 2^{\cal P})$ where $v: 2^{\cal P} \longrightarrow \mathbb{R}$ .
We provide an axiomatic characterization of the two-step flow associated with a two-step method inspired by the configuration value studied in Albizuri et al. \cite{Albizuri2006a} for TU-games with coalition configuration of the form $(N, v, {\cal P}, 2^{\cal P})$ where $v: 2^N \longrightarrow \mathbb{R}$ and $v (\emptyset) = 0$. We propose two axioms for a flow on $\Gamma_{2^{\cal P}}$. 
These axioms have the same flavor of the two axioms of anonymity used by Owen \cite{Owen1977} to characterize the Owen value. \\
 
\noindent {\bf Notation}: For a permutation $\sigma: A \longrightarrow A$ over the elements of a finite set $A$, the notation $\sigma (S)$, $S \subseteq A$, denotes the image of $S$ by $\sigma$. \\

\noindent {\bf Intracoalitional anonymity}. A flow $\Lambda$ on $\Gamma_{2^{\cal P}}$ satisfies intracoalitional anonymity if for each coalition profile ${\cal K} \in  2^{\cal P}$, each $q \in M$
each $i \in P_q$, each coalition $K_q \subseteq P_q \setminus i$ and each permutation $\sigma_q : P_q \longrightarrow P_q $ of the elements in $P_q$ the following principle of invariance on the flow holds:
$$\Lambda \bigl((K_q,  {\cal K}_{-q}),  (K_q \cup i,  {\cal K}_{-q})\bigr) = \Lambda \bigl((\sigma_q (K_q),  {\cal K}_{-q})),  (\sigma_q (K_q \cup i),  {\cal K}_{-q})\bigr).$$

\noindent {\bf Coalitional anonymity}. A flow $\Lambda$ on $\Gamma_{2^{\cal P}}$ satisfies coalitional anonymity if for each $q\in M$, each $S \subseteq M$ such that $S \ni q$
and each permutation $\sigma : M \longrightarrow M $ of the elements of $M$, the following principle of invariance of the aggregate flow holds:
$$\sum_{i \in P_q} \Lambda ({\cal K}_{S\setminus q},  {\cal K}_{S, \{i\})}) = \sum_{i \in P_{\sigma(q)}} \Lambda ({\cal K}_{\sigma (S\setminus q)},  {\cal K}_{\sigma(S), \{i\}}).$$

 \begin{theorem} \label{AZ-theo}
Among the flow methods on ${\cal G}_{N, {\cal P},  2^{\cal P}}$,  there is a unique flow method $\Phi^{AZ}$ whose associated flow $\Lambda_{AZ}$ on $\Gamma_{2^{{\cal P}}} $  satisfies Null flow for non-relevant directed edges,   Intracoalitional anonymity and Coalitional anonymity.  Furthermore, $\Phi^{AZ}$ is a two-step flow method such that,
for each relevant directed edge, its associated  flow $\Lambda_{AZ}$ is defined as:
 $$ \Lambda_{AZ} ({\cal K}_{S, K_q}, {\cal K}_{S, K_q \cup i}) = \Lambda^M_{AZ} ({\cal R}_{S \setminus q},  {\cal R}_{S})  \Lambda^q_{AZ} (K_q, K_q \cup i),$$
where
 $$ \Lambda^M_{AZ} ({\cal R}_{S \setminus q},  {\cal R}_{S}) = \frac{(m - s)! (s - 1) !}{m !} \,\, \mbox { and } \,\, \Lambda^q_{AZ} (K_q, K_q \cup i) = \frac{(p_q - k - 1)! k!}{p_q !},$$
and $s= |S|$,  $k = |K_q|$. The flow passing through non-relevant edges is null.
The corresponding two-step flow method $\Phi^{AZ}$ on ${\cal G}_{N, {\cal P},  2^{\cal P}}$ is given by:
$$\forall i \in N, \quad \Phi^{AZ}_i (N, v, {\cal P}, 2^{\cal P}) =$$
\begin{equation} \label{notre-AZ}
 \sum_ {q \in M: \atop i \in P_q } \sum_{S \subseteq M: \atop S \ni q} \sum_{K_q \in 2^{P_q}: \atop K_q \not \ni i}\frac{(m - s) ! (s - 1)!}{m !}  \frac{(p_q - k - 1)! k!}{p_q !}
\bigl (v({\cal K}_{S, K_q \cup i}) - v({\cal K}_{S, K_q} )\bigr). 
\end{equation}
 \end{theorem} 
 
{ \it Proof} By definition,  $\Lambda_{AZ}$ satisfies Null flow for non-relevant directed edges,   Intracoalitional anonymity and Coalitional anonymity. Reciprocally, assume that there is a flow $\Lambda$ on $\Gamma_{2^{{\cal P} }} $ that satisfies these axioms.
 As  $\Lambda$  satisfies  Null flow for non-relevant directed edges, Lemma \ref{Null-flow-M} applies.
Consider the unitary flow  $\Lambda^M$ passing through the directed edges of the hypercube $\Gamma_{(2^{\cal P})^M}$ as in (\ref{flowMprod}) in the statement of Lemma \ref{Null-flow-M}:
$$ \forall S \subseteq M \setminus \emptyset, \forall q \in S, \quad 
\Lambda^M ({\cal R}_{S \setminus q}, {\cal R}_S) = \sum_{i \in P_q} \Lambda_{} ({\cal K}_{S\setminus q}, {\cal K}_{S, \{i\} }). 
$$
Coalitional anonymity implies that the flow $\Lambda^M ({\cal R}_{S \setminus q}, {\cal R}_S)$ uniquely depends on the size $s$ of $S$.
Pick any $s \in \{1, \ldots, m\}$ and consider the following cut $(F_{s - 1}, F_{s})$  in the hypercube $\Gamma_{(2^{\cal P})^M}$:
$$F_{s - 1} = \{{\cal R}_S :  | S| \leq s-1 \}  \mbox { and } F_{s } = \{{\cal R}_S :  | S| \geq s \}.$$  
Since the flow  $\Lambda^M$ is unitary, the value of the cut is
$$1 = \sum_{({\cal R}_S, {\cal R}_{S' }) \in F_{s - 1} \times F_{s }} \Lambda^M ({\cal R}_{S}, {\cal R}_{S'}) = \sum_{S \in 2^M: \atop |S| = s} \sum_{q \in S}   \Lambda^M({\cal R}_{S\setminus q}{\cal R}_{S}).$$
As $\Lambda^M ({\cal R}_{S \setminus q}, {\cal R}_S)$ uniquely depends on the size $s$ of $S$, all the flows in the above sum are equal, and the number of  directed edges
$({\cal R}_{S\setminus q}{\cal R}_{S})$ where $| S| = s$ is given by
$${m \choose s} s,$$
so that the flow $\Lambda^M ({\cal R}_{S \setminus q}, {\cal R}_S)$ of each such directed edge 
is uniquely determined:
 $$\Lambda^M ({\cal R}_{S \setminus q}, {\cal R}_S) = \frac{ (m - s) ! (s - 1) !}{m !} = \Lambda_{AZ}^M ({\cal R}_{S \setminus q}, {\cal R}_S),$$
 as desired.
 
 Next, by Point 1 of Lemma \ref{Null-flow-M}, the unitary flow $\Lambda$ on $\Gamma_{2^{\cal P}}$ induces a unitary flow on the subdigraph 
 $({(2^{\cal P})}^{S\setminus q, S},  E_{(2^{\cal P})^{S\setminus q, S}})$ and the value of this flow is $\Lambda^M ({\cal R}_{S \setminus q}, {\cal R}_S)$. Therefore, for $k \in \{0, \ldots, p_q - 1\}$,
 consider the cut  $(F^S_{k },  F^S_{k +1 })$ of the sudigraph $({(2^{\cal P})}^{S\setminus q, S},  E_{(2^{\cal P})^{S\setminus q, S}})$ given by
 $$F^S_{k} = \bigl\{ {\cal K}_{ S,  K_q} : |K_q| \leq k \bigr\}  \mbox { and } F^S_{k +1 } = \bigl \{ {\cal K}_{ S,  K_q} : |K_q| > k \bigr\},$$ one obtains 
 $$\sum_{K_q \in 2^{P_q}: \atop |K_q| = k} \sum_{i \in P_q \setminus K_p} \Lambda ( {\cal K}_{ S,  K_q}, {\cal K}_{ S,  K_q \cup i})    =  \frac{(m - s) ! (s - 1) ! }{m !}.$$
Note that
$$ ({\cal K}_{S,  K_q}, {\cal K}_{S,  K_q \cup i}) =  (K_q, ({\cal K}_{ S})_{ -q}, K_q \cup i, ({\cal K}_{ S})_{ -q}).$$ 
By Intracoalitional anonymity, $\Lambda ({\cal K}_{ S,  K_q}, {\cal K}_{ S,  K_q \cup i})$ only depends on the size $k$ of $K_q$. It follows that all the terms of the above double sum are equal and the number of directed edges $ (K_q, ({\cal K}_{ S})_{ -q}, K_q \cup i, ({\cal K}_{ S})_{ -q})$ is given by
$${p_q \choose k} (p_q - k). $$
 It follows that
$$  \Lambda({\cal K}_{ S,  K_q}, {\cal K}_{ S,  K_q \cup i}) = \frac{(m - s) ! (s - 1) ! }{m !} \frac{ (p_q - k - 1) ! k !}{p_q !}.$$
To conclude the proof, set
 $$\Lambda_{AZ}^q (K_q,  K_{q} \cup i )=  \frac{ (p_q - k - 1) ! k !}{p_q !},$$
from which one obtains that  $\Lambda_{AZ}^q$ is a unitary flow on the hypercube of dimension $p_q$.
 \qed
  
 The two step-method defined in  (\ref{notre-AZ}) can be viewed as the extension of the configuration value introduced by Albizuri et al. \cite{Albizuri2006a} from TU-games with a coalition configuration to games in ${\cal G}_{N, {\cal P},  2^{\cal P}}$ where the coalition profile function is defined on $2^{{\cal P}}$ and not on $2^N$.  A TU-game  with a coalition configuration is given by a triple $(N, v, {\cal P})$ where $v: 2^N \longrightarrow \mathbb{R}$, with the convention $v (\emptyset) = 0$. The {\bf configuration value} $\Phi^C$ is defined as:
 $$\forall i \in N, \quad \Phi^{C}_i (N, v, {\cal P}) =  $$
 \begin{equation} \label{leur-AZ}
 \!\!\!  \sum_ {q \in M: \atop i \in P_q } \!\!\! \sum_{S \subseteq M: \atop S \ni q, P_{S \setminus q} \not \ni i }  \!\!\!  \sum_{K_q \in 2^{P_q}: \atop K_q \not \ni i}\frac{(m - s) ! (s - 1)!}{m !}  \frac{(p_q - k - 1)! k!}{p_q !}
\biggl (v\bigl((K_q \cup i) \cup P_{S \setminus q} \bigr) - v\bigl( K_q \cup P_{S \setminus q} \bigr)\biggr),
 \end{equation}
where 
$$P_{S \setminus q} = \cup_{r \in S \setminus q} P_r.$$
The main difference between  $\Phi^{C}$ and  $\Phi^{AZ}$ is that in the former if
$ i \in P_{S \setminus q}$,
then agent's $i$ contribution to coalition $K_q \cup P_{S \setminus q}$ is null:
$$v\bigl((K_q \cup i) \cup P_{S \setminus q} \bigr) - v\bigl( K_q \cup P_{S \setminus q} \bigr) = 0.$$
This is not the case for the two-step method  $\Phi^{AZ}$ in  (\ref{notre-AZ}) where the contribution
$$v({\cal K}_{S, K_q \cup i}) - v({\cal K}_{S, K_q} ),$$
is not necessary null whenever $i \in P_{S \setminus q}$. 
The two-step method $\Phi^{AZ}$ considers instead that  if $i$ has already contributed to an element $P_r$, it can continue to contribute in another element $P_q$ of the coalition configuration to which it belongs. That is the reason why $\Phi^{AZ}$ computes agent $i$'s contribution for each $S \subseteq M$ such that $q \in S$.  In other words, by considering coalition profiles and not only coalitions, agent $i$ is viewed as an active agent in $P_q$ even if it has already joined other elements $P_r$ to which it belongs. 

The question that naturally emerges from the above discussion and which is addressed in the next section is the possibility of obtaining a flow method for games $ (N, v , {\cal P}, {\cal F}) $
where $v$ is a classical coalition function $v : 2^N \longrightarrow \mathbb{R}$, where $v (\emptyset) = 0$, from a flow method on ${\cal G}_{N, {\cal P},  {\cal F}}$.

 \section{From coalition profiles to coalitions } \label{coalitionp-coalition}
 
For each coalition profile ${\cal K} = (K_q)_{q \in M} \in {\cal F}$, we define the coalition $ u({\cal K}) \in 2^N$ as
 $$ u({\cal K}) = \bigcup_{q \in M} K_q. $$
Reciprocally, a coalition $ R \in 2^N$ is {\bf reachable} from ${\cal F}$ if
 $$\exists {\cal K} \in {\cal F}: \quad   R =   u({\cal K}).$$
Let ${\cal F}^0 \subseteq 2^N$ be the set of reachable coalitions from ${\cal F}$. 
Note that $(N, {\cal F}^0)$  is a normal set system. Indeed, $\emptyset$ is reachable from $\emptyset _{M}$ and $N$ is reachable from ${\cal P}$ since, by assumption,
$$\cup_{q \in M} P_q = N.$$
From  ${\cal F}^0$ consider the {\bf directed graph} $\Gamma_{{\cal F}^0}^* = ({\cal F}^0, E_{{\cal F}^0}^*)$ where
$$\forall R, R' \in {\cal F}^0, R \not = R',    \quad (R, R') \in E_{{\cal F}^0}^* \mbox { if } \exists ({\cal K}, {\cal K}') \in E_{\cal F}: R = u({\cal K}), R' = u({\cal K}'). $$
In words, $(R, R')$ is a directed edge of $\Gamma_{{\cal F}^0}^*,$ if $R$ and $R'$ are reachable from the endpoints of some directed edge of $\Gamma_{{\cal F}}$.

 \begin{example}
Consider again  Example \ref{graph1}. The set of reachable coalitions ${\cal F}^0$  from the product set system ${\cal F}$ is:
\begin{eqnarray}
\biggl \{ \emptyset, \{1\},  \{2, 3\},   \{3, 4\},& & \!\!\!\{1, 2, 3\}, \{1, 3, 4\},   \{2,3, 4\},   \{3, 4, 5\}, \{1, 2, 3, 4\},  \nonumber \\
& &  \{1, 3, 4, 5\},  \{2, 3, 4, 5\},  \{1, 2, 3, 4, 5\}   \!   \biggr\}. \nonumber
\end{eqnarray} 
For instance, coalition $\{2, 3, 4, 5\} \in {\cal F}^0$ is obtained from ${\cal K} =  (\{2, 3\}, \{3, 4, 5\}) \in {\cal F}$.
Figure \ref{R-coalition} represents the digraph $\Gamma_{{\cal F}^0}^* = ({\cal F}^0, E_{{\cal F}^0}^*)$.
For instance, the directed edge $(\{1, 3, 4\}, \{1, 3, 4, 5\}) \in E_{{\cal F}^0}^*$ is obtained from the directed edge $({\cal K}, {\cal K}') = 
((\{1\}, \{3, 4\}), (\{1\}, \{3, 4, 5\}))$ of the product digraph $\Gamma_{\cal F}$ represented in Figure $\ref{fig2-1}$. Indeed,
$u ({\cal K}) = \{1, 3, 4\}$ and  $u ({\cal K}') = \{1, 3, 4, 5\}.$

\begin{figure}[h]
\begin{center}
\begin{tikzpicture} [every node/.style={scale = 0.60}, scale = 0.60]
  \node (A) at (0,0) {$ \emptyset$};
  \node (B) at (5,0) {$\{2, 3\},$};
  \node (C) at (2.5, 1.5) {$\{1\}$};
  \node (D) at (7.5, 1.5) {$\{1, 2, 3\}$};

  \draw[->] (A) -- (B);
  \draw[->] (B) -- (D);
  \draw [->] (C) -- (D);
  \draw [->] (A) -- (C);

 \node (E) at (0, 3) {$ \{3, 4\}$};
 \node (F) at (5, 3) {$ \{2, 3, 4\}$};
\node (G) at (2.5 , 4.5) {$\{1, 3, 4\}$};
\node (H) at (7.5, 4.5) {$ \{1, 2, 3, 4\}$};  
  
   \draw[->] (E) -- (F);
  \draw[->] (E) -- (G);
  \draw [->] (G) -- (H);
  \draw [->] (F) -- (H);
  
  \node (I) at (0, 6) {$\{3, 4, 5\}$};
\node (J) at (5, 6) { \{2, 3, 4, 5\}};
\node (K) at (2.5, 7.5) {$ \{1, 3, 4, 5\}$};
\node (L) at (7.5, 7.5) { $\{1, 2, 3, 4, 5\}$};

   \draw [->] (I) -- (J);
 \draw [->] (I) -- (K);
 \draw [->] (J) -- (L);
 \draw [->] (K) -- (L);  

\draw [->] (A) -- (E);
\draw [->] (E) -- (I);

  \draw [->] (B) -- (F);
  \draw [->] (F) -- (J);
  
  \draw [->] (C) -- (G);
  \draw [->] (G) -- (K);  
  
    \draw [->] (D) -- (H);
  \draw [->] (H) -- (L);

  \end{tikzpicture}
\end{center}
\caption{The digraph of reachable coalitions $\Gamma_{ {\cal F}^{0} }^* = ({\cal F}^{0}, E_{{\cal F}^{0}}^*) $ obtained from ${\cal F}$.}
\label{R-coalition}
\end{figure}
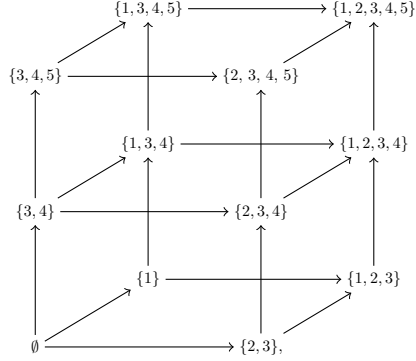
\end{example}
 
 \begin{remark} \label{digraphs} The digraph of reachable coalitions $\Gamma_{ {\cal F}^{0} }^* = ({\cal F}^{0}, E_{{\cal F}^{0}}^*) $ of Figure \ref{R-coalition} is also the digraph
$ \Gamma_{ {\cal F}^{0} } = ({\cal F}^{0}, E_{{\cal F}^{0}})$ of the covering relation
 of the partial ordered set $({\cal F}^0, \subseteq)$. This correspondence between these two digraphs does not always hold. In general, both digraphs are distinct. To see this, examine the  following example.
 
Assume that $N = \{1, \ldots, 5\}$ and $M = \{1, 2\} $ where $P_1 = \{1, 2, 3, 4\}$, $P_2 = \{1, 5\}$ and
 ${\cal F}_1 = \{\emptyset, \{1, 2\}, \{2, 3\}, \{1, 2, 3, 4\}\}$,
${\cal F}_2 = \{\emptyset,  \{1, 5\}\}$. The product digraph $\Gamma_{\cal F}  = ({\cal F},  E_{\cal F})$  is the directed hypercube of dimension 2. Consider the coalition $R = \{1, 2, 5\}$ and $R' = \{1, 2, 3, 5 \}$ of $2^N$. Coalition $R$ is reachable
from ${\cal K} = (\{1, 2\}, \{1, 5\})$ and  $R'$ is reachable from ${\cal K}' = (\{2, 3\}, \{1, 5\})$. Obviously, 
$R' $ covers $R$ in the partial ordered set $({\cal F}^0, \subseteq)$. However, it is easy to see that $(R, R')$ is not a directed edge of $\Gamma_{ {\cal F}^{0} }^*$.
Reciprocally, $ (\{1, 5\}, \{1, 2, 3, 5\} )$ is a directed edge of $\Gamma_{ {\cal F}^{0} }^*$ because coalition $\{1, 5\}$ is reachable from ${\cal K} = (\emptyset, \{1, 5\})$,
coalition $\{1, 2, 3, 5\}$ is reachable from ${\cal K}' = (\{2, 3\}, \{1, 5\})$ and $ ({\cal K}, {\cal K}') \in E_{\cal F}$. Nevertheless, $(\{1, 5\}, \{1, 2, 3, 5\} )$ is not a directed edge  
of $\Gamma_{{\cal F}^0} = ({\cal F}^0, E_{{\cal F}^0})$.
Figure \ref{fig-Ex-product} represents $\Gamma_{{\cal F}^0}^*$ and the directed graph $\Gamma_{{\cal F}^0} = ({\cal F}^0, E_{{\cal F}^0})$ of the covering relation of $({\cal F}^0, \subseteq)$.

\begin{figure}[h]
\begin{center}
\begin{tikzpicture} [every node/.style={scale = 0.60}, scale = 0.60]
  \node (A) at (0,0) {$\emptyset$};
  \node (B) at (5,0) {$\{1, 2\}$};
  \node (C) at (2.5, 1.5) {$\{2, 3\}$};
  \node (D) at (7.5, 1.5) {$\{1, 2, 3, 4\}$};

  \draw[->] (A) -- (B);
  \draw[->] (B) -- (D);
  \draw [->] (C) -- (D);
  \draw [->] (A) -- (C);

 \node (E) at (0, 3) {$\{1, 5\}$};
 \node (F) at (5, 3) {$ \{1,2, 5\}$};
\node (G) at (2.5 , 4.5) {$\{1, 2, 3, 5\}$};
\node (H) at (7.5, 4.5) {$\{1, 2, 3, 4, 5\}$};  
  
   \draw[->] (E) -- (F);
  \draw[->] (E) -- (G);
  \draw [->] (G) -- (H);
  \draw [->] (F) -- (H);
  
  \draw [->] (A) -- (E);
 \draw [->] (B) -- (F);
   \draw [->] (C) -- (G);
\draw [->] (D) -- (H);
  
  
  \node (A') at (11,0) {$\emptyset$};
  \node (B') at (16,0) {$\{1, 2\}$};
  \node (C') at (13.5, 1.5) {$\{2, 3\}$};
  \node (D') at (18.5, 1.5) {$\{1, 2, 3, 4\}$};

  \draw[->] (A') -- (B');
  \draw[->] (B') -- (D');
  \draw [->] (C') -- (D');
  \draw [->] (A') -- (C');

 \node (E') at (11, 3) {$\{1, 5\}$};
 \node (F') at (16, 3) {$ \{1,2, 5\}$};
\node (G') at (13.5 , 4.5) {$\{1, 2, 3, 5\}$};
\node (H') at (18.5, 4.5) {$\{1, 2, 3, 4, 5\}$};  
  
  \draw [->]  (F')-- (G');
   \draw[->] (E') -- (F');
  \draw [->] (G') -- (H');
  \draw [->] (F') -- (H');

\draw [->] (A') -- (E');
 \draw [->] (B') -- (F');
 \draw [->] (C') -- (G');
 \draw [->] (D') -- (H');

  \end{tikzpicture}
\end{center}
\caption{To the left, $\Gamma_{{\cal F}^0}^*$, to the right  the directed graph $\Gamma_{{\cal F}^0}$ of the covering relation of $({\cal F}^0, \subseteq)$.}
\label{fig-Ex-product}
\end{figure}
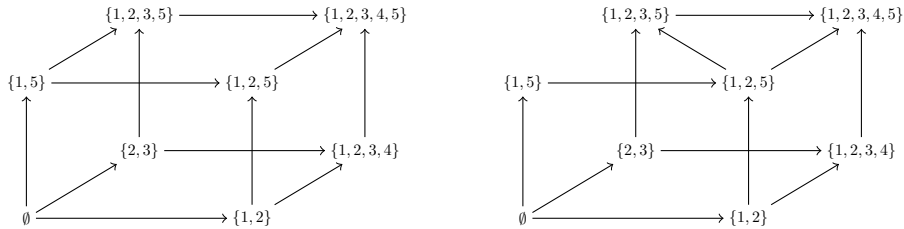

 \end{remark}

 As in previous sections, assume that $N$, ${\cal P}$ and ${\cal F}$ are fixed once and for all
and define  ${\cal G}_{N, {\cal F}^0, {\cal P}}$ as the set of TU-games
$(N, v, {\cal F}^0, {\cal P})$ with restricted cooperation ${\cal F}^0 \subseteq 2^N$  
 and coalition configuration ${\cal P}$ where $v: {\cal F}^0 \longrightarrow \mathbb{R}$ and $v( \emptyset) = 0$.
 For any game $(N, v, {\cal F}^0, {\cal P}) \in {\cal G}_{N, {\cal F}^0, {\cal P}}$ we associate the game $(N, v^*, {\cal F}, {\cal P}) \in {\cal G}_{N, {\cal F}, {\cal P}}$ defined as: 
$$\forall {\cal K} \in {\cal F}, \quad v^*({\cal K}) = v( u({\cal K})).$$
This implies that
$$\forall {\cal K}, {\cal K}' \in {\cal F}, \quad  \bigl [ u({\cal K}) =  u({\cal K}') \bigr ]\Longrightarrow  \bigl [v^*({\cal K}) = v^*({\cal K}') = v( u({\cal K}))\bigr ].$$

 For any value $\Phi$ on ${\cal G}_{N, {\cal F}, {\cal P}}$, we define the value 
$\phi $ on $ {\cal G}_{N, {\cal F}^0, {\cal P}}$ as follows:
 $$\forall (N, v, {\cal F}^0, {\cal P}) \in {\cal G}_{N, {\cal F}^0, {\cal P}}, \quad \phi (N, v, {\cal F}^0, {\cal P})  = \Phi  (N, v^*, {\cal F}, {\cal P}).$$ 
We write, $\phi \triangleleft \Phi.$

The following result provides a sufficient condition on the product digraph $\Gamma_{{\cal F}} = ({\cal F}, E_{\cal F})$ such that if $\Phi$ is a flow method on ${\cal G}_{N, {\cal F}, {\cal P}}$, then
$\phi \triangleleft \Phi$ is a flow method on $ {\cal G}_{N, {\cal F}^0, {\cal P}}$ whose corresponding flow $\Lambda^{\phi}$ passes through the directed edges of $\Gamma_{{\cal F}^0}^*$.

\begin{remark}
Flow methods on $ {\cal G}_{N, {\cal F}^0, {\cal P}}$ are defined in analogous way to  flow methods on  ${\cal G}_{N, {\cal F}, {\cal P}}$. 
The main difference with Definition 2.4 of flow methods for TU-games with restricted cooperation in Aguilera et al. \cite{Aguilera2010} is that they consider a flow on the digraph of the covering relation with respect to set inclusion of the elements of the set system. As showed in Remark \ref{digraphs}, $\Gamma_{{\cal F}^0}^*$ may differ from $\Gamma_{{\cal F}^0}$.
To the contrary, we consider a flow passing through the directed edges of $\Gamma_{{\cal F}^0}^*$ and not to the directed edges of $\Gamma_{{\cal F}^0}$.

\end{remark}

\begin{theorem} \label{final}
Assume that for each directed edge $({\cal K}, {\cal K}') \in E_{\cal F}$, 
\begin{enumerate}
\item either $u({\cal K}) =  u({\cal K}'),$
\item or $u({\cal K}') \setminus  u({\cal K}) = K_q' \setminus K_q $ and $q$ is the only index in $M$ such that $K_q' \not = K_q$.
\end{enumerate}
Then, for any flow method $\Phi$  on ${\cal G}_{N, {\cal F}, {\cal P}}$,   $\phi \triangleleft \Phi$ is a flow method  on ${\cal G}_{N, {\cal F}^0, {\cal P}}$ whose associated flow
passes through the directed of edges of $\Gamma_{{\cal F}^0}^*$ .
\end{theorem} 

{\it Proof} Assume that $\Gamma_{{\cal F}}$ is as the statement of Theorem \ref{final}. Pick any flow method $\Phi$  on ${\cal G}_{N, {\cal F}, {\cal P}}$
and consider  $\phi \triangleleft \Phi$ on ${\cal G}_{N, {\cal F}^0, {\cal P}}$. Then,
$$ \forall (N, v, {\cal F}^0, {\cal P}) \in {\cal G}_{N, {\cal F}^0, {\cal P}}, \forall i \in N, \quad \phi_i (N, v, {\cal F}^0, {\cal P})  =   \Phi_i (N, v^*, {\cal F}, {\cal P}),      $$
that is, by definition of a flow method  and $v^*$,
\begin{eqnarray}
 \phi_i (N, v, {\cal F}^0, {\cal P}) &= & \sum_{({\cal K}, {\cal K}') \in E_{\cal F}^i} \lambda_{i}^{\Phi}({\cal K}, {\cal K}') \bigl(v^*( {\cal K}') - v^*({\cal K}) \bigr) \nonumber \\
 & = & \sum_{(R, R') \in E_{{\cal F}^0}^*} \sum_{({\cal K}, {\cal K}') \in E_{\cal F}^i: \atop u ({\cal K}) = R, u({\cal K}') = R' } \lambda_{i}^{\Phi}({\cal K}, {\cal K}') \bigl(v (R') - v(R) \bigr). \nonumber 
\end{eqnarray}
By definition, $({\cal K}, {\cal K}') \in E_{\cal F}^i$ if  $i \in Q _{({\cal K}, {\cal K}')}$. Thus,
whenever  $u ({\cal K}) \not = u({\cal K}')$, Point 2 of the statement of Theorem \ref{final} implies that $ R' \setminus R =  Q _{({\cal K}, {\cal K}')}$, which leads to
$$ \phi_i (N, v, {\cal F}^0, {\cal P}) =  \sum_{(R, R') \in E^*_{{\cal F}^0}: \atop i \in R' \setminus R} \sum_{({\cal K}, {\cal K}') \in E_{\cal F}: \atop u ({\cal K}) = R, u({\cal K}') = R' } \lambda_{i}^{\Phi}({\cal K}, {\cal K}') \bigl(v (R') - v(R) \bigr).$$
By setting,
$$\forall (R, R') \in E_{{\cal F}^0}^*: i \in R' \setminus R, \quad \lambda_i^\phi (R, R') =  \sum_{({\cal K}, {\cal K}') \in E_{\cal F}: \atop u ({\cal K}) = R, u({\cal K}') = R' }\lambda_{i}^{\Phi}({\cal K}, {\cal K}'),$$
we get
$$ \phi_i (N, v, {\cal F}^0, {\cal P}) =  \sum_{(R, R') \in E_{{\cal F}^0}^*: \atop i \in R' \setminus R} \lambda_i^\phi (R, R')\bigl(v (R') - v(R) \bigr),$$
from which one concludes that $\phi$ is a marginalist value on ${\cal G}_{N, {\cal F}^0, {\cal P}}$. Next, by Efficiency of $\Phi$, one obtains
$$\sum_{i \in N}  \phi_i (N, v, {\cal F}^0, {\cal P}) =   \sum_{i \in N}\Phi_i (N, v^*, {\cal F}, {\cal P}) = v^*({ \cal P} ) = v (u ( { \cal P}) ) = v(N).$$
Thus, $\phi$ is Efficient.  Next, proceeding as in the proof of Theorem \ref{flow-Eff} but for the digraph $\Gamma_{{\cal F}^0}^*$ and for a coalition profile function $v: {\cal F}^0 \longrightarrow \mathbb{R}$ one obtains that $\phi$ is a flow method on  ${\cal G}_{N, {\cal F}^0, {\cal P}}$ whose associated flow $\Lambda^{\phi}: E_{{\cal F}^0}^* \longrightarrow \mathbb{R}$ is given by
$$\forall (R, R') \in   E_{{\cal F}^0}^*, \quad \Lambda^{\phi}(R, R') = \sum_{i \in R' \setminus R} \lambda_i^\phi (R, R').$$
This concludes the proof.
\qed

 \begin{remark}  \label{conclude}
1. Conditions 1 and 2 of the statement of Theorem \ref{final} are not satisfied for the product digraph $\Gamma_{\cal F}$ of Example \ref{graph1}.
 To see this, consider the directed edge $( (\{2, 3\}, \emptyset), (\{2, 3\}, \{3, 4\} ) ) \in E_{{\cal F}}$. Then, $u (\{2, 3\}, \emptyset ) = \{2, 3\}$,
$ u (\{2, 3\}, \{3, 4\}) = \{2, 3, 4\} $,  and   $\{2, 3, 4\}  \setminus \{2, 3\} = \{4\}$ which is different from $ \{3, 4\} \setminus \emptyset$.

2. Conditions 1 and 2 of the statement of Theorem \ref{final} are satisfied when the product digraph $\Gamma_{\cal F}$ results from any coalition configuration ${\cal P} = (P_q)_{q \in M}$ of $N$ and regular set systems $(P_q, {\cal F}_q)$, $q \in M$. Indeed, in such a case $Q_{({\cal K}, {\cal K}')}$ is a singleton whatever $({\cal K}, {\cal K}') \in E_{\cal F}$. Note also that if each $(P_q, {\cal F}_q)$ is regular, then $\Gamma_{{\cal F}^0}^* = \Gamma_{{\cal F}^0}$.

3. Conditions 1 and 2 of the statement of Theorem \ref{final} are satisfied when the product digraph $\Gamma_{\cal F}$ results from a partition ${\cal P} = (P_q)_{q \in M}$ of $N$ whatever ${\cal F}_q$, $q \in M$. In this case,   $\Gamma_{{\cal F}^0}^* = \Gamma_{{\cal F}^0}$ since there is a one-to-one mapping between the coalitions in ${\cal F}^0$ and the coalition profiles in ${\cal F}$.   \end{remark}
 
From Points 2 and 3 of Remark \ref{conclude}, one obtains the following corollary from Theorem \ref{final}.

\begin{corollary} \label{conclude-2}
Assume that one of the following conditions holds:
\begin{enumerate}
\item  The coalition configuration ${\cal P}$ is a partition of $N$;
\item Each set system $(P_q, {\cal F}_q)$, $q \in M$, is regular.
\end{enumerate}
Then, for any flow method $\Phi$  on ${\cal G}_{N, {\cal F}, {\cal P}}$,   $\phi \triangleleft \Phi$ is a flow method  on ${\cal G}_{N, {\cal F}^0, {\cal P}}$ and the associated flow passes through the directed edges of  $\Gamma_{{\cal F}^0}$. 
\end{corollary}

From Point  1 of Corollary \ref{conclude-2},  the flow method $\phi^{AZ} \triangleleft \Phi^{AZ}$ where $\Phi^{AZ}$ is defined in (\ref{notre-AZ}) and characterized in Theorem \ref{AZ-theo},  generalizes the Owen value (1977) from TU-games with a partition to TU-games with a partition and restricted cooperation. In a similar way, Point  2 of Corollary \ref{conclude-2}  leads to the conclusion that the flow method $\phi^{AZ} \triangleleft \Phi^{AZ}$ generalizes the configuration value of Albizuri et al. \cite{Albizuri2006a} from TU-games with a coalition configuration to TU-games with a coalition configuration and restricted cooperation by regular set systems.  Note that regular set systems contain convex geometries or normal augmenting systems, among others. However, normal accessible union stable networks or intersecting stable networks introduced in Algaba et al. in \cite{A-Brink-D-2018} are also interesting examples of set systems. Indeed, although they satisfy the property of accessibility and one point extension, respectively (both of them satisfied by regular set systems), they are not necessarily regular set systems.

 \section{Conclusion} \label{concl}
We introduce generalized coalition configurations in which the cooperation in each element of the coalition configuration is not necessarily total but takes into account the possible incompatibilities among the agents. This leads to a variety of novel situations depending on the particular set system in each element of the coalition configuration. First, we provide an axiomatic characterization of flow methods for cooperative games with generalized coalition configuration. Next, we look at a value construction procedure inspired by Owen's two-step procedure 
for TU-games with coalition structure (Owen \cite{Owen1977}). This two-step procedure leads to a flow method whose flow can be decomposed into two flows. Such decomposable flows are characterized by two axioms for flows. Finally, we investigate the possibility of deriving a flow method for TU-games with generalized coalition configuration  from a flow method for our class of cooperative games where the unit of cooperation is a coalition profile and not a coalition.

{\bf Acknowledgements.} We would like to thank the participants of the SING 19 conference, Elena Molis, Adriana Navarro-Ramos and the participants of the seminar of the Department of Economic Theory and Economic History at the University of Granada. 
This work is part of the R\&D\&I project grant PID2022-137211NB-100, funded by MCIN/ AEI/10.13039/501100011033/ and by ``ERDF A way of making Europe/EU''. Financial support from the Universit\'e Jean Monnet, Saint-Etienne, within the program ``Math\'ematiques de la d\'ecision pour l'ing\'enierie physique et sociale'' (MODMAD) is acknowledged. Eric R\'emila and Philippe Solal would like to thank the hospitality received during their stay in the Universidad de Sevilla (IMUS), Escuela  Superior de Ingenieros.\\

   \end{document}